\documentclass[a4paper,11pt]{article}
\pdfoutput=1 

\usepackage{jheppub} 

\usepackage[T1]{fontenc} 
\usepackage{slashed}
\usepackage[colorlinks=true,linktocpage=true]{hyperref}
\usepackage{cleveref}
\usepackage[numbers,sort&compress]{natbib}
\usepackage{mathtools}
\usepackage{xspace}

\newcommand{\gev}{\ensuremath{\,\text{GeV}}\xspace}
\newcommand{\tev}{\ensuremath{\,\text{TeV}}\xspace}

\crefname{figure}{Fig.{}}{Figs.{}}
\crefname{section}{Sec.{}}{Secs.{}}
\crefname{table}{Tab.{}}{Tabs.{}}
\crefname{equation}{Eq.{}}{Eqs.{}}

\title{\boldmath Dilution of dark matter relic density in singlet extension models}

\author[a,b]{Yang Xiao,}
\author[a,b]{Jin Min Yang,}
\author[c,a]{Yang Zhang}

\affiliation[a]{CAS Key Laboratory of Theoretical Physics, Institute of Theoretical Physics, Chinese Academy of Sciences, Beijing 100190, P. R. China}
\affiliation[b]{School of Physical Sciences, University of Chinese Academy of Sciences,  Beijing 100049, P. R. China}
\affiliation[c]{School of Physics, Zhengzhou University, ZhengZhou 450001, P. R. China}

\emailAdd{xiaoyang@itp.ac.cn}
\emailAdd{jmyang@itp.ac.cn}
\emailAdd{zhangyangphy@zzu.edu.cn}

\abstract{
We study the dilution of dark matter~(DM) relic density caused by the electroweak first-order phase transition~(FOPT) in the singlet extension models, including the singlet extension of the standard model~(xSM), of the two-Higgs-doublet model~(2HDM+S) and the next-to-minimal supersymmetric standard model~(NMSSM). We find that in these models the entropy released by the strong electroweak FOPT can dilute the DM density to 1/3 at most. Nevertheless, in the xSM and NMSSM where the singlet field configure is relevant to the phase transition temperature, the strong FOPT always happens before the DM freeze-out, making the dilution effect negligible for the current DM density. 
We derive an analytical upper bound on the freeze-out temperature and a numerical lower bound on nucleation temperature in the xSM.
On the other hand, in the 2HDM+S where the DM freeze-out temperature is independent of FOPT, the dilution may salvage some parameter space excluded by excessive DM relic density or by DM direct detections.
}

\begin{document} 
\maketitle
\flushbottom

\section{Introduction}
\label{sec:intro}

The electroweak symmetry is broken in the present Universe, but is supposed to be restored in the early Universe~\cite{KIRZHNITS1972471}. 
In the Standard Model~(SM), the electroweak symmetry-breaking occurs through a crossover transition~\cite{kajantie1997non}, 
while it can be first-order phase transition~(FOPT) in presence of Beyond-the-SM (BSM) physics. 
Studying of FOPT has been of heightened interest, because it could have provided the conditions needed for explanation of the baryon asymmetry of the Universe~(BAU)~\cite{RevModPhys.71.1463,Cline:2006ts,10.1088/978-1-6817-4457-5,Morrissey:2012db} 
and generated detectable gravitational wave~(GW)~\cite{PhysRevLett.69.2026,Kamionkowski:1993fg}. 

The BAU is characterized by the baryon-to-entropy density ratio $Y_B \equiv n_{B}/s$ and the most precise measurement is given by Planck~\cite{Planck:2015fie} 
\begin{equation}
    Y_B = 
    8.65\pm0.09 \times 10^{-11},
\end{equation}
which is consistent with the value obtained from measurements of primordial abundance for light elements~\cite{Workman:2022ynf}. The BSM physics to explain it must satisfy the so-called Sakharov criteria: C- and CP-violation, baryon number violation and departure from equilibrium~\cite{sakharov1998violation}. Many mechanisms have been proposed following such criteria~\cite{PhysRevLett.92.061303,PhysRevLett.93.201301,PhysRevD.43.984,PhysRevD.56.6155, PhysRevD.64.063511,RevModPhys.71.1463,1982AZh,alberghi2007radion}. The electroweak baryogenesis is one of the most attractive scenarios and has been widely studied ~\cite{RevModPhys.71.1463, PhysRevD.45.2685, PhysRevD.53.4578, Ghosh_2021, Prokopec_2019, Baldes_2019, Wang:2022dkz}, which requires an electroweak FOPT. The transition occurs when the bubbles of the symmetry-broken phase nucleate in plasma of the symmetry-restored phase. More baryons are produced than antibaryons in the regions around the expanding bubble walls. To avoid washing out the created baryons, the transition has to be strongly first-order. 

The collisions, sound waves and turbulence from expanding bubbles of the broken phase can generate detectable GWs~\cite{Maggiore:1999vm,Weir:2017wfa,Alanne:2019bsm}. The observation of GW signal~\cite{LIGOScientific:2016aoc} has opened up a new window to probe BSM, especially in the situation that the LHC searches for BSM have merely given null results so far. The GW generated through strong FOPT is within coverage of future GW detectors, such as the Laser Interferometer Space Antenna~(LISA)~\cite{2017arXiv170200786A} and the Taiji program~\cite{10.1093/nsr/nwx116}.

The other by-product of strong FOPT is the change of Dark Matter~(DM) density. When the Universe evolved from the symmetric phase to the broken phase, there was an entropy injection and latent heat release, which could dilute the DM density. The DM relic density in the present Universe has been measured precisely by astrophysical and cosmological experiments: $\Omega_{\rm DM} h^2 = 0.120\pm0.001$~\cite{Planck:2018vyg}. The DM models, such as the Weakly Interacting Massive Particle (WIMP)\cite{PhysRevD.100.115050, PhysRevD.82.055026, PhysRevD.80.055012}, right hand neutrino \cite{PhysRevD.101.095005, PhysRevD.79.033010, PhysRevD.81.085032, PhysRevD.90.013007} and axion\cite{PhysRevLett.120.211602, PhysRevLett.108.061304, PhysRevD.91.065014}, should give the correct relic density. Assuming a multi-component DM, the proportions of different DM candidates also affect the interpretation of DM direct and indirect searching results \cite{PhysRevD.103.063028, PhysRevD.100.015040, PhysRevLett.119.181301}. 

For those DM models in which the FOPT occurs after DM freeze-out in the evolution history of the Universe, it is mandatory to take the dilution of DM density into consideration.
The dilution effect can be induced by the entropy injection, which may come from heavy states decaying to the thermal bath~\cite{Dine_1996, Banks_1994} or the electroweak FOPT~\cite{megevand2004first, megevand2008supercooling,wainwright2009impact,chung2011probing,azatov2021dark,azatov2022ultra}. 
Refs.~\cite{megevand2004first, megevand2008supercooling} first showed some general information on the amounts of supercooling and reheating as well as the duration of the phase transition using thermodynamic method. 
Then in a model-independent way Ref.~\cite{wainwright2009impact} found that the FOPT may dilute the thermal relic abundance of DM if the decouple process finished before the electroweak transition. Further, Ref.~\cite{chung2011probing} systematically studied various imprints of the phase transitions on the relic abundance of TeV-scale DM. The above studies showed that the dilution factor, $ d=(a_f/a_i)^3$ with $a_i(a_f)$ being the  scale factor of the Universe at the beginning (ending) of the phase transition, can be as large as 50. It will distinctly affect the DM signal expectations in relevant experiments.  However, the above studies focused on the dilution effect by using the toy-model-like effective potential such as adding large degree of freedom from hidden sector. In the studies of realistic models emphasizing DM aspect~\cite{bian2018thermally,bian2019two,Han:2020ekm, blinov2015electroweak,mcdonald2012secluded}, the dilution effect has been neglected by a rough estimation. 

In this work, we aim to study in detail the dilution of DM relic density caused by FOPT in realistic models. Therefore we focus on singlet extension models, including the scalar singlet extension of the SM~(xSM), of two-Higgs-doublet model~(2HDM+S) and the next-to-minimal supersymmetric Standard Model~(NMSSM). These models are well motivated and popular in studying electroweak phase transition and DM. The introduced singlet can both trigger FOPT and provide a good DM candidate. We will calculate the magnitude of the dilution factor and find out the conditions of successful dilution of DM relic density in these models.

This paper is organized as follows. In \cref{sec:mechanism} we introduce the dilution mechanism caused by FOPT and deduce formulas for calculating the dilution factor. We describe the models and present results for xSM in \cref{sec:xsm}, for 2HDM+S in \cref{sec:2HDM+S}, and for NMSSM in \cref{sec:nmssm}.

\section{Dilution of DM density by first-order phase transition}
\label{sec:mechanism}

\subsection{Electroweak first-order phase transition}

The amount of dilution on DM density depends upon the relative entropy injected into the broken phase during the FOPT. Two special temperatures of the FOPT, the critical temperature ($T_c$) and nucleation temperature ($T_n$) defined in the following, are the vital features in determining the dilution factor. 

In the hot and radiation-dominated early Universe, the electroweak symmetry is restored, i.e. the minimum of Higgs potential locates at origin $v_{\rm origin}$. As the temperature of the Universe drops, a second minimum $v_{\rm broken}$ away from the origin develops with higher free energy. 
With the Universe further cooling, the symmetry-broken minimum becomes degenerate with the origin, which gives the definition of critical temperature  
\begin{equation}
    V_{\rm eff} (v_{\rm origin};T_c) = V_{\rm eff} (v_{\rm broken};T_c).
\end{equation}
Then, with temperature falling below $T_c$, some regions of the symmetric plasma tunnel to the deeper broken minimum and nucleate bubbles. 
Most of the bubbles are too small to grow and they just collapse, because the energy difference between the two vacuums is not large enough to overcome the surface tension of the bubble walls.  
As the Universe further cools, the nucleation rate of large bubbles increases dramatically. 
The phase transition begins once the probability to nucleate a supercritical bubble in one Hubble volume is of order one, at the so-called nucleation temperature $T_n$.

Quantitatively, the tunneling probability per unit time per unit volume can be roughly estimated as~\cite{linde1983decay}
\begin{equation}
    \Gamma \sim T^4~e^{-\frac{S_{E}}{T}},
\end{equation}
where $S_{E}$ is the three-dimensional Euclidean action given by
\begin{equation}
    S_{E} =  4 \pi \int^{+\infty}_{0}r^2{\rm d}r~\left[\frac{1}{2}\left(\frac{\partial \phi}{\partial r}\right)^{2} + V_{\rm eff}(\phi;T)\right].
\end{equation}
The bubble configuration $\phi(r)$ in the integral is fixed from the corresponding Euclidean equation of motion
\begin{equation}
    \frac{{\rm d}^2 \phi}{{\rm d} r^2} + \frac{2}{r}\frac{{\rm d} \phi}{{\rm d}r} = \frac{\partial V_{\rm eff}(\phi;T)}{\partial \phi},
\end{equation}
subjecting to the boundary conditions $\lim \limits_{r \to \infty} \phi(r) = 0 $ and ${\rm d} \phi/{\rm d} r|_{r=0}=0$ (see \cite{rubakov2009classical} for details). The $\mathcal{O}(1) $ probability of nucleating a supercritical bubble in one Hubble volume is expressed as 
\begin{equation}
    \int^{+\infty}_{T_{n}} \frac{{\rm d} T}{T}(\frac{2\zeta M_{PI}}{T})^4 e^{-\frac{S_{E}}{T}} = \mathcal{O}(1),
\end{equation}
where $M_{PL}$ is the reduced Planck mass, $\zeta = \frac{1}{4 \pi}\sqrt{\frac{45}{\pi g^{*}}}$ with $g^{*}$ being the effective number of relativistic degrees of freedom \cite{quiros1998finite}. From this definition, we can get an estimated formula for $T_{n}$
\begin{equation} \label{eq: S/T}
    \frac{S_{E}}{T_{n}} \sim \mathcal{O}(130-140).
\end{equation}

\subsection{Dilution mechanism via entropy injection}

There are two situations for the dilution of DM density due to entropy injection. In the first situation, the transition temperature $T_n$ is near the critical temperature $T_c$, and the supercooling process is negligible. 
The system is almost in equilibrium and thus the total entropy $(\sim a^3 s)$ is conserved, where $s$ indicates the entropy density and $a$ is the scale factor of the Universe. With entropy injection, the density changes as $s_- = (a_i/a_f)^3 ~ s_{+}$, where the subscripts ${+}$ and ${-}$ denote the high-temperature symmetric phase and low-temperature broken phase, and the subscripts $f$ and $i$ indicate the
beginning and ending of the phase transition. Therefore we can get the dilution factor
\begin{equation}  \label{eq: entropy scale}
    d \equiv \left(\frac{a_{f}}{a_i}\right)^{3} = \frac{s_{+}(T_c)}{s_{-}(T_n)},
\end{equation}
for the transition in this situation.

In the second situation, where the transition is strongly first-order and  $T_{n}$ is consequently much smaller than $T_{c}$, the equilibrium condition is broken at a stage of the transition. For convenience, we can divide the evolution of transition into supercooling stage, reheating stage and phase coexistence stage.

In the supercooling stage, the high temperature phase always dominates the Universe, so the total entropy is conserved. Similar to the first situation, we have 
\begin{equation} 
\label{eq:supercooling}
    \left(\frac{a_{i}}{a_{m}}\right)^{3} \simeq \frac{s_{+}(T_{n})}{s_{+}(T_{c})},
\end{equation}
where $a_{m}$ is the scale factor at a temperature near $T_n$, corresponding to the end of supercooling. 

Then the latent heat is released and reheats the Universe. If the duration of the reheating stage is short compared to the expansion rate, the energy density $\rho$, instead of the total entropy, is conserved~\cite{wainwright2009impact}. When the Universe is reheated to a temperature close to $T_c$, it reaches a phase coexistence stage, and its energy density can be expressed as
\begin{equation} 
\label{eq: f_define}
    \rho_-(T_c) = f \rho_{-}(T_{c}) + (1-f)\rho_{+}(T_{c}), 
\end{equation}
where $f$ is the volumetric fraction of the plasma of the low-temperature phase. With energy density conservation, the energy density at the beginning of the reheating stage is given as
\begin{equation} 
\begin{aligned} 
    \rho_+(T_n) & = \rho_-(T_c) \\
    & =  \rho_{+}(T_{c}) - f[\rho_{+}(T_{c}) - \rho_{-}(T_{c})].
\end{aligned}
\end{equation}
Then we can get the fraction
\begin{equation} \label{eq: f_cal}
    f = \frac{\rho_{+}(T_{c})-\rho_{+}(T_{n})}{L},
\end{equation}
where $L = \rho_{+}(T_{c}) - \rho_{-}(T_{c}) $ is the latent heat.

During the third stage, which also happens quickly, the total entropy is again conserved. The total entropy is approximately $a_m^3 ~[(1-f)s_{+}(T_c) + f s_{-}(T_c)] $ at the beginning and $a_f^3 ~s_{-}(T_c) $ at the ending of this stage. Therefore we see
\begin{equation}
    \left(\frac{a_{f}}{a_{m}}\right)^3 = \frac{1-f\Delta s / s_{+}(T_{c})}{1 - \Delta s/s_{+}(T_{c})},
\end{equation}
where $\Delta s = s_{+}(T_c) - s_{-}(T_c)$.
Finally, combined with \cref{eq:supercooling}, we get the total expansion of dilution factor
\begin{equation} \label{eq: supercolling dilution factor}
  d \equiv \left(\frac{a_{f}}{a_{i}}\right)^3 = (\frac{1-f\Delta s / s_{+}(T_{c})}{1 - \Delta s/s_{+}(T_{c})})(\frac{s_{+}(T_{c})}{s_{+}(T_{n})}).
\end{equation}
for the second situation.

\subsection{DM relic density}
\label{sec:dm}
\begin{figure}[t]
\centering 
\includegraphics[width=0.6\textwidth]{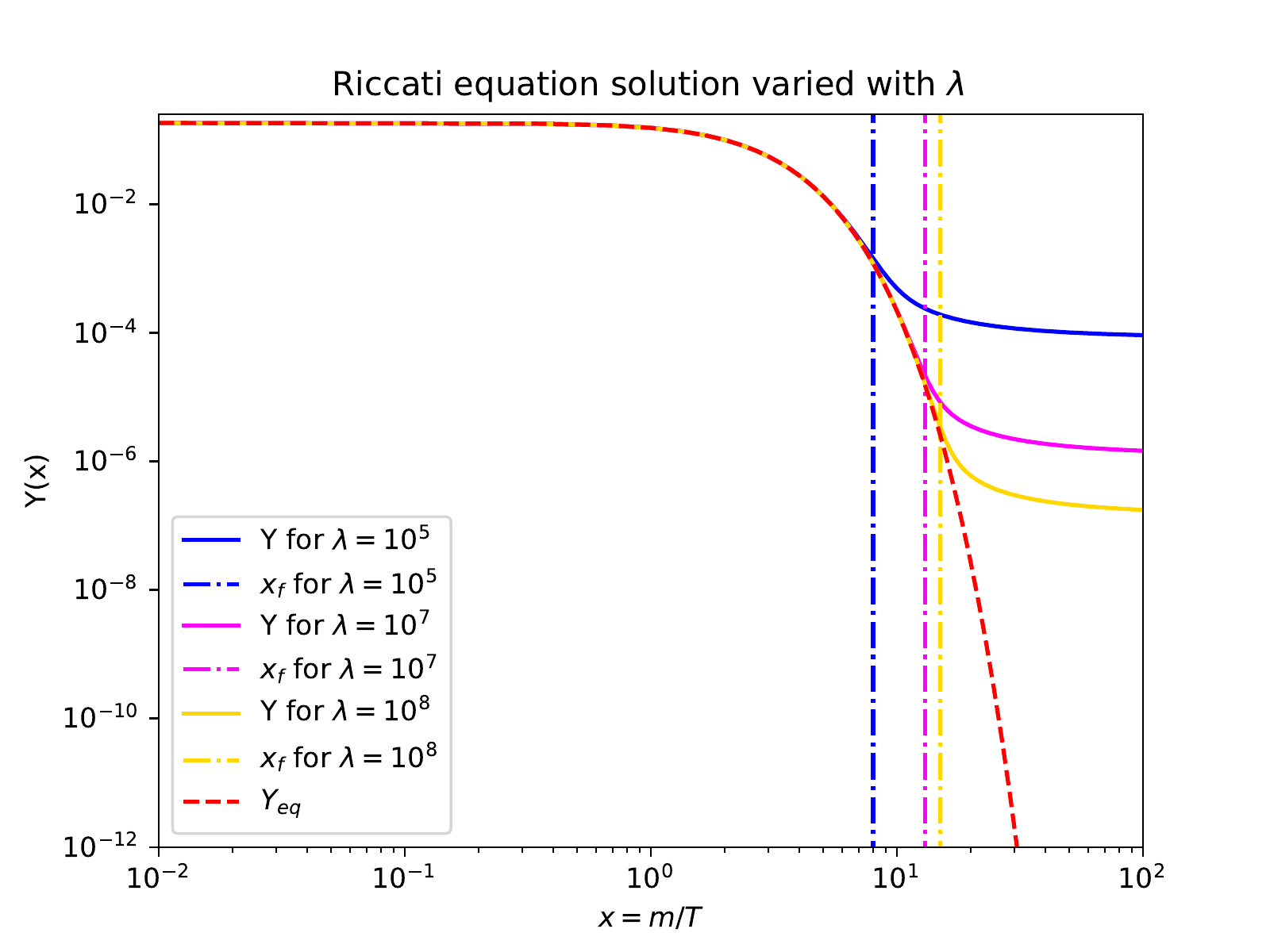}
\caption{The solution of the Riccati equation with varying $\lambda$. The red dashed line shows the $Y_{\rm eq}$ at the thermal equilibrium. The blue, pink and yellow solid line are Riccati solution when $\lambda = 10^5, 10^7, 10^8$, respectively.}
\label{fig:Riccati}
\end{figure}

The relic density of freeze-out DM is calculated through solving the Boltzmann transport equation \cite{PhysRevLett.39.165}
\begin{equation}
    \frac{\mathrm{d} n_{\rm DM}}{\mathrm{d} t} = -3Hn_{\rm DM} - \left<v_{\rm rel}\sigma\right>(n_{\rm DM}^2-n_{\rm DM,eq}^{2}),
\end{equation}
where $n_{\rm DM}$ is the DM number density, $n_{\rm DM,eq}$ is the DM number density at thermal equilibrium with the rest of the universe, $H$ is the Hubble rate and $\left<v_{\rm rel}\sigma\right>$ is the relative velocity of the annihilating DM particles times the thermally averaged self-annihilation cross-section. This equation states that the change of the DM density comes from two parts: (1) the dilution effect due to Hubble expansion, (2) the particle reactions including DM production and DM annihilation. Trading $n_{\rm DM}$ and $t$ with $Y=n_{\rm DM}/T^3$ and $x=m_{\rm DM}/T$ respectively, the Boltzmann equation becomes the Riccati equation
\begin{equation} \label{eq: riccati equation}
    \frac{\mathrm{d} Y}{\mathrm{d} x} = -\frac{\lambda}{x}(Y^2-Y_{\rm eq}^2),
\end{equation}
where 
\begin{equation}
    \lambda = \sqrt{\frac{45}{4 \pi^3 g_{*}}}m_{\rm DM}M_{p}\left<v_{\rm rel}\sigma\right>.
\end{equation}
By dimensional analysis, the magnitude is estimated as $\lambda = 1.59 \times 10^8 {\mathcal{F}}/{g_{*}} ({m_{\rm DM}}/{1 \gev})^3$, where $\mathcal{F}$ is a fudge factor to take account of the number of annihilation channels and
the details of the interaction responsible for the annihilation~\cite{weinberg2008cosmology, young2017survey}. Treating $\lambda$ as a constant number, we can solve the Riccati equation numerically with certain  boundary conditions, such as shown in \cref{fig:Riccati}. 

At high temperature, we have $x = m_{\rm DM}/T \ll 1$ and $\lambda/x\gg 1$, so there is a strong negative feedback effect on $Y$. Any deviation with respect to $Y_{\rm eq}$ will lead to an opposite derivative in the right side of \cref{eq: riccati equation}, and thus $Y$ remains at $Y_{\rm eq}$ at high temperatures. When the universe cools down below the DM mass, $Y_{\rm eq}$ will decrease exponentially as $e^{-x}$, so does $Y$. Finally, when the temperature is much lower than the DM mass, i.e. below the so-called freeze-out temperature $T_{f} = m_{\rm DM}/x_{f}$, ${\lambda}/{x}$ is so small that the DM deviates from the thermal equilibrium and $Y$ turns to a constant, which corresponds to the DM relic density. 

The dilution effect caused by FOPT can alter the curves of $Y(x)$ displayed in Fig.~\ref{fig:Riccati}. If the freeze-out temperature is higher than the transition temperature, there will be a drop at $x=m_{\rm DM}/T_n>x_f$ in the flat region, whose magnitude is decided by the dilution factor in Eq.~\ref{eq: dilution approx}. Thus, the final DM density can be calculated using the transitional method and then multiplying by the dilution factor $d$. However, in the numerical tools such as \texttt{MicrOmega}~\cite{belanger2002micromegas} and \texttt{MadDM}~\cite{Ambrogi:2018jqj}, the DM mass and couplings in the Boltzmann equation are set to the values with spontaneous symmetry breaking at zero-temperature, which is fine when $T_n>T_f \sim 0~\gev$. In our case, the DM mass should be calculated in the electroweak symmetry restored vacuum with thermal corrections. Fortunately, we see in Fig.~\ref{fig:Riccati} that the freeze-out temperature $T_f$ does not change too much with $\lambda$ varying 
in a large range. Therefore, in the following, we still estimate the DM relic density using  \texttt{MicrOmega}, and leave the sophisticated calculation for future work. 
On the other hand, if $T_n>T_f$, the drop will happen on the left side of $x_f$. Owing to the strong negative feedback effect on $Y$ in \cref{eq: riccati equation}, the value of $x_f$ is barely changed, as well as the DM relic density.

\section{Singlet extension of the SM}
\label{sec:xsm}

\subsection{Effective potential}

Firstly, we consider xSM, the simplest scalar extension of the SM,  which includes an extra $Z_{2}$ symmetric real scalar singlet field $S$.  The scalar potential is given by \begin{equation}\label{eq:xsm_tree}
    V_{0}(H,S) = -\mu_{H}^{2}H^{\dagger}H + \lambda_{H}(H^{\dagger}H)^{2} - \frac{\mu_{S}^2}{2}S^2 + \frac{\lambda_{S}}{4}S^4 + \frac{\lambda_{HS}}{2}H^{\dagger}HS^2.
\end{equation}
The Higgs doublet can be parameterized as
\begin{equation}\label{eq: parameterized H}
H = 	{\left[\begin{array}{c}
	G^{+}\\ \frac{ h + iG^{0}}{\sqrt{2}}
	\end{array}\right]},
\end{equation}
where $G^{\pm, 0}$ indicate the Goldstone bosons and $h$ stands for the SM Higgs boson. We parameterize the $S$ field as $S = s$.  The background field configurations $h$ and $s$ have vacuum expectation values (VEVs) of $(v_h,v_s)$ at zero-temperature. Substituting them into Eq~(\ref{eq:xsm_tree}), the tree level tadpole conditions are 
\begin{equation}
\label{eq: tadpole conditions_tree}
\begin{aligned}
   \left<\frac{\partial V_{0}}{\partial h}\right> &= v_{h} \left(-\mu_{H}^2 + \lambda_{H}v_{h}^2 + \frac{\lambda_{HS}}{2} v_{s}^2\right) = 0,\\
    \left<\frac{\partial V_{0}}{\partial s}\right> &= v_{s} \left(-\mu_{S}^2 + \lambda_{S}v_{s}^2 + \frac{\lambda_{HS}}{2} v_{h}^2 \right) = 0,
\end{aligned}
\end{equation}
where $\left<...\right>$ represents that the quantity in between the angled brackets is evaluated in the vacuum $(h=v_h,s=v_s)$. 
Among the solutions of \cref{eq: tadpole conditions_tree}, we take the $Z_2$ invariant one, $v_h = \mu_H^2/\lambda_H$ and $v_s=0$. 
The scalar masses are obtained by diagonalizing the squared mass matrix
evaluated at the VEV,
\begin{equation}
\label{eq: mass matrix}
\begin{aligned}
M^2 &= {\left(                 
  \begin{array}{cc}   
\left<\frac{\partial^2 V_{0}}{\partial^2 h^2}\right> & \left<\frac{\partial^2 V_{0}}{\partial h \partial s}\right>\\
\left<\frac{\partial^2 V_{0}}{\partial h \partial s}\right> &  \left<\frac{\partial^2 V_{0}}{\partial^2 s}\right>
  \end{array}
\right)}\\
&= {\left(                 
  \begin{array}{cc}   
3v_{h}\lambda_{H} - \mu_{H}^2 + \frac{1}{2}\lambda_{HS}v_{s}^2 & \lambda_{HS}v_{h}v_{s}\\
\lambda_{HS}v_{h}v_{s} &  3v_{s}\lambda_{S} - \mu_{S}^2 + \frac{1}{2}\lambda_{HS}v_{h}^2
  \end{array}
\right)}.\\
\end{aligned}
\end{equation}
Since $v_{s} =0$, the off-diagonal elements of the mass matrix are eliminated, i.e. no mixing between the Higgs and singlet fields. Owing to this, the singlet scalar particle does not interact with any other particles except the Higgs boson, which consequently is a viable DM candidate. 

The relic density of this DM candidate is mainly determined by Higgs funnel annihilation. The cross section for annihilation into SM particles except Higgs is~\cite{Cline:2013gha}
\begin{equation}
    \sigma v_{\rm rel} = \frac{2\lambda_{hs}^2v_h^2 }{\sqrt{\mathbf{s}}} \frac{\Gamma_h(\sqrt{\mathbf{s}})}{(\mathbf{s}-m_h^2)^2+m_h^2\Gamma_h^2(m_h)} ,
\end{equation}
where $\Gamma_h(m_h^*)$ is the full Higgs boson width as a function of invariant mass. Its thermal average is given as~\cite{Gondolo:1990dk}
\begin{equation}
    \left< \sigma v_{\rm rel} \right> = \int_{4m_S^2}^{\infty} \frac{\mathbf{s}\sqrt{\mathbf{s}-4m_S^2}K_1(\sqrt{\mathbf{s}/T})\sigma v_{\rm rel}}{16T m_S^4 K_2^2 (m_S/T)} \mathrm{d} \mathbf{s},
\end{equation}
where $K_1$ and $K_2$ are the second kind modified Bessel functions. As the annihilation is via the $s$-wave, the relic density can be estimated as 
\begin{equation}
    \Omega_{S} h^2 \sim \frac{3\times 10^{-27}\mathrm{cm}^3/\mathrm{s}}{\left< \sigma v_{\rm rel} \right>} \times d, 
\end{equation}
if the freeze-out temperature is higher than the transition temperature.

To calculate the dilution factor of phase transition, we need the effective potential with loop corrections
\begin{equation} \label{eq: V_eff}
V_{\rm eff}(h, s; T) = V_{0}(h, s) + V_{\rm CW}(h, s) + V_{\rm CT}(h, s) +  V_{1T}(h, s; T) + V_{\rm ring}(h, s; T), 
\end{equation}
where $V_{\rm CW}$, $V_{\rm CT}$, $V_{1T}$ and  $V_{\rm ring}$ are the one-loop Coleman-Weinberg potential, the corresponding counter term, the one-loop thermal correction and the resummed daisy correction, respectively. 

We choose the OS-like scheme and the Landau gauge to avoid introducing dependence on renormalization scale~\cite{quiros1998finite}. The one-loop zero-temperature correction takes a form~\cite{PhysRevD.45.2685}
\begin{equation} \label{eq: V_CW}
\begin{aligned}
V_{1}(h, s)  =& V_{\rm CW}(h, s) + V_{\rm CT}(h, s) \\
 =& \sum_{i} (-1)^{s_{i}} \frac{g_{i}}{64 \pi^2}  \left\{ m_{i}^4(h, s)\left[\log \frac{m_{i}^2(h, s)}{m_{i}^2(v_{h}, v_{s})} -\frac{3}{2} \right] + 2m_{i}^2(h, s)m_{i}^2(v_{h}, v_{s}) \right\},
\end{aligned}
\end{equation}
where $i \in \{H,~S,~W^{\pm},~Z,~\gamma,~t$\}, $s_i$ is the spin of particle $i$, and $g_i$ is the number of degrees of freedom, 
\begin{equation}
g_{H} = 1, ~~ g_{S} = 1, ~~ g_{W^{\pm}} = 6, ~~ g_{Z} = 3, ~~ g_{\gamma}=3, ~~ g_{t} = 12.  
\end{equation}
$m_i^2(h, s)$ stands for the squared tree-level background-field-dependent masses,
\begin{equation}
\begin{aligned}
m_{H}^2 &= - \mu_{H}^2 + 3\lambda_{H}h  + \frac{1}{2}\lambda_{HS}s^2 \\
m_{S}^2 &= - \mu_{S}^2 + 3\lambda_{S}s  + \frac{1}{2}\lambda_{HS}h^2 \\
m_{W^{\pm}} & = \frac{1}{4} g_{SU(2)_L}^2 h^2, ~~~~~
m_{Z} = \frac{1}{4} \left(g_{SU(2)_L}^2+g_{U(1)_Y}^2\right) h^2 \\
m_{\gamma}^2 &=0, ~~~~~~~~~~~~~~~~~~~
m_{t}  = \frac{1}{2} y_t^2h^2.\\
\end{aligned}    
\end{equation}
We neglect the contributions of light fermions. Goldstone bosons are not included because that second derivative of $V_{\rm CW}$ is logarithmic divergent at VEV of zero-temperature, originating from $\left.m_{G^{\pm,0}}\right|_{h=v_h,s=0}=0$ and $\left. \partial m_{G^{\pm,0}}/\partial h \right|_{h=v_h,s=0}\neq0$~\cite{elias2014taming}. The impact of fixing the Goldstone catastrophe, as well as choosing other renormalization scheme, can be found in Ref.~\cite{braathen2016avoiding,Athron:2022jyi}.

In the OS-like scheme, the position of VEV at zero-temperature and masses at VEV are not affected by loop corrections, so the renormalization conditions are imposed as
\begin{equation}
\begin{aligned}
\left< \frac{\partial (V_{0} + V_{1})}{\partial h}\right> &= \left< \frac{\partial V_{0}}{\partial h}\right> = 0, \\
\left<\frac{\partial^2 (V_{0} + V_{1})}{\partial^2 h}\right> &= \left<\frac{\partial^2 V_{0}}{\partial^2 h}\right> = m_H^2, ~~
\left<\frac{\partial^2 (V_{0} + V_{1})}{\partial^2 s}\right> = \left<\frac{\partial^2 V_{0}}{\partial^2 s}\right> = m_S^2.
\end{aligned}    
\end{equation}
Thus, with $v_h=246\gev$ and $m_H = 125\gev$, we can fix $\mu_H$ with \cref{eq: tadpole conditions_tree,eq: mass matrix}. We choose $m_S$, $\lambda_S$ and $\lambda_{HS}$ as input parameters of the model. 

The one-loop finite temperature correction is deduced from the finite-temperature field theory~\cite{quiros1998finite}
\begin{equation} \label{eq: one-loop effective potential at finite temperature}
    V_{1T}(h, s) = \frac{T^{4}}{2 \pi^2} \left[\sum_{B} g_{B}J_{B}\left(\frac{m_{B}(h, s)}{T}\right) + \sum_{F} g_{F}J_{F}\left(\frac{m_{F}(h, s)}{T}\right)\right],
\end{equation}
where $J_{B}$, $J_{F}$ are the relevant thermal distribution functions for the bosonic and fermionic contributions, respectively,
\begin{equation} \label{eq: J_BJ_F}
    J_{B,F}(x) = \pm \int_{0}^{+\infty} {\rm d}y ~y^2 \log \left(1\mp e^{-\sqrt{x^2 + y^2}}\right).
\end{equation}
In the high-temperature limit, they can be expanded as
\begin{equation}
\label{eq:high-temperature JBF}
\begin{aligned}
J_{B}(x) &= -\frac{\pi ^4}{45} + \frac{\pi^2}{12}x^2 -\frac{\pi}{6}x^3 - \frac{x^4}{32}\log\frac{x^2}{a_b},\\
J_{F}(x) &= -\frac{7\pi^4}{360} + \frac{\pi^2}{24}x^2 + \frac{x^4}{32}\log\frac{x^2}{a_f},
\end{aligned}
\end{equation}
where $\log(a_{b})=\frac{3}{2}-2\gamma + 2\log(4\pi)$, $\log({a_{f})=\frac{3}{2}-2\gamma+2\log(\pi)}$, with $\gamma$ being the Euler constant~\cite{PhysRevD.45.2685}. 

Owning to the zero-mode contribution, the multi-loop diagrams will dominate if the temperature is high enough, which makes the perturbation condition broken~\cite{curtin2018thermal, senaha2020symmetry}. To make the expansion reliable, the dominant thermal pieces must be resummed. Here we adopt the Parwani method~\cite{Parwani:1991gq} by replacing the tree-level masses $m_{i}^2$ in \cref{eq: V_CW,eq: one-loop effective potential at finite temperature}) through the thermal masses $m_{i}^2(T) = m_{i}^2 + d_{i}T^2$, where $d_{i}T^2$ is the leading contribution in temperature to the one-loop thermal mass, where
\begin{equation} \label{eq: deby mass correction}
\begin{aligned}
d^{L}_{W^{\pm,3}}&=\frac{11}{6}g_{SU(2)_L}^2, ~~~~~~
d^{T}_{W^{\pm,3}}=0, ~~~~~~
d_{B}^{L}=\frac{11}{6}g_{U(1)_Y}^{2},\\
d_{HH} &= \frac{3g_{SU(2)_L}^2}{16} + \frac{1}{16}g_{U(1)_Y}^2+\frac{1}{2}\lambda_{H}+\frac{1}{4}y_{t}^2+\frac{1}{24}\lambda_{HS},\\
d_{SS} &= \frac{1}{4}\lambda_{S} + \frac{1}{6}\lambda_{HS}.
\end{aligned}
\end{equation}

\subsection{Scan results}

We performed a scan in the following ranges,
\begin{equation} \label{eq: scan_area}
    10\gev \le m_{S} \le 1\tev, ~~0 \le \lambda_{HS} \le 10, ~~0 \le \lambda_{S} \le 1.
\end{equation}
The upper bounds on $\lambda_S$ and $\lambda_{HS}$ are set because of perturbation limits. We will see later that the upper bound on $m_S$ is large enough for achieving strong FOPT. \texttt{Cosmotransition}~\cite{WAINWRIGHT20122006} and \texttt{PhaseTracer}~\cite{Athron:2020sbe} are used to calculate the physical quantities related to the phase transition, and \texttt{MicrOmega} is used to get the freeze-out temperature $T_{f}$ and DM observables. In the following, we study samples that can achieve successfully FOPT in the thermal history of the Universe.
The bound on DM relic density from Planck~\cite{Planck:2015fie} and limits from DM direct detection XENON1T~\cite{PhysRevLett.119.181301} are discussed later taking the dilution effect into consideration. For other possible constraints on xSM, see Refs.~\cite{Athron:2018ipf}. 

\subsubsection{Dilution factor}

\begin{figure}[t]
\centering 
\includegraphics[width=0.48\textwidth]{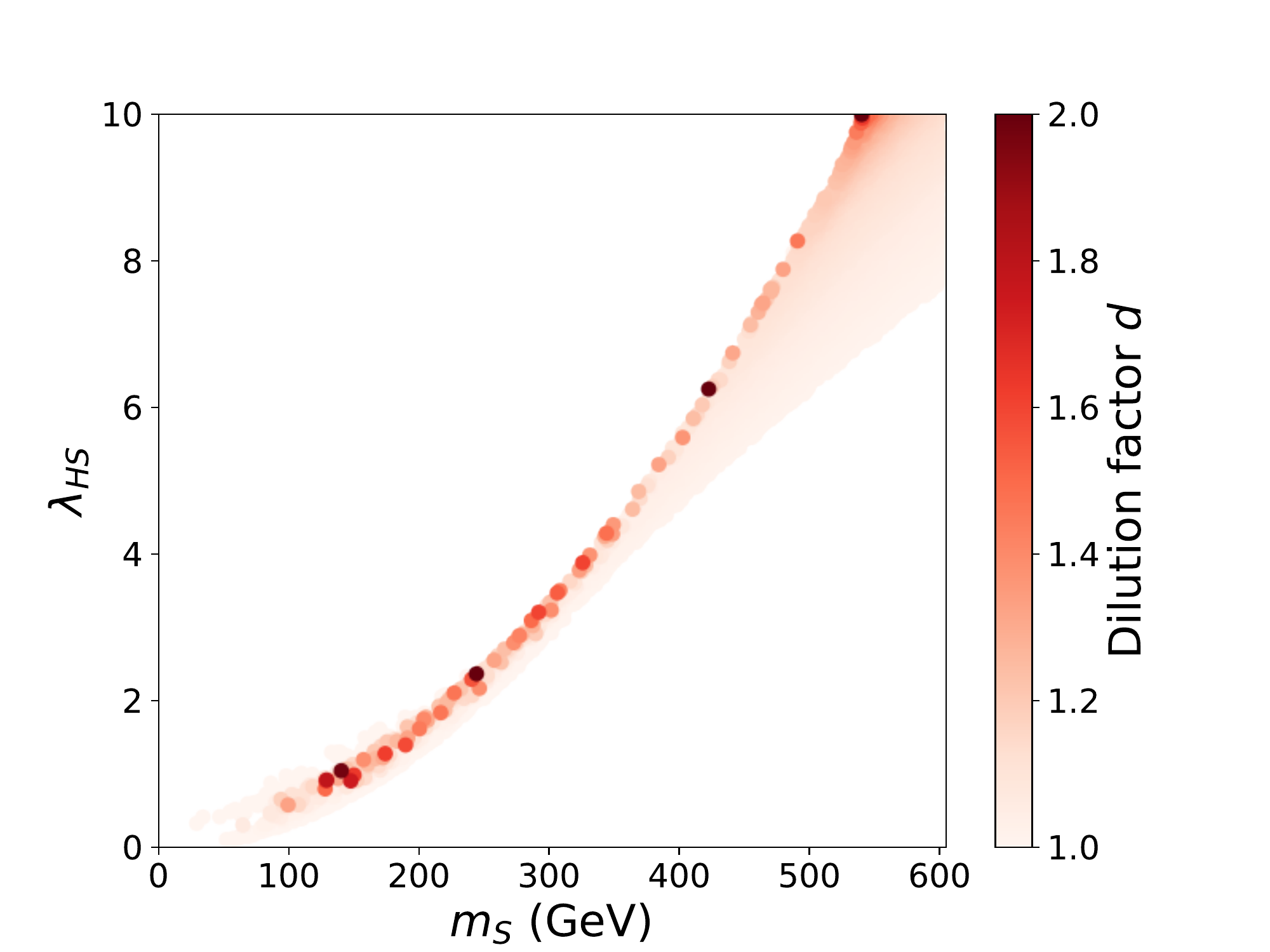}
\hfill
\includegraphics[width=0.48\textwidth]{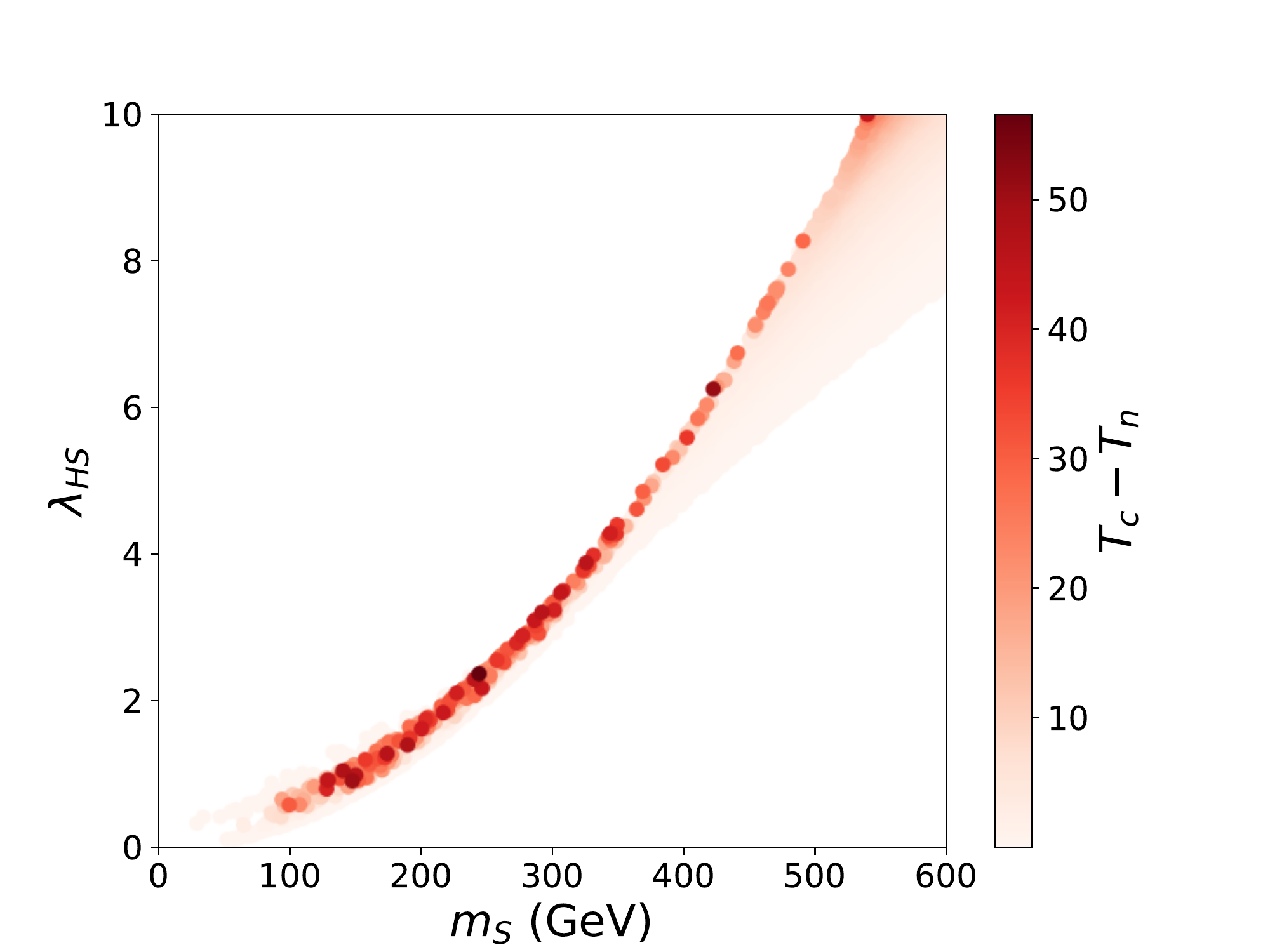}
\hfill
\includegraphics[width=0.48\textwidth]{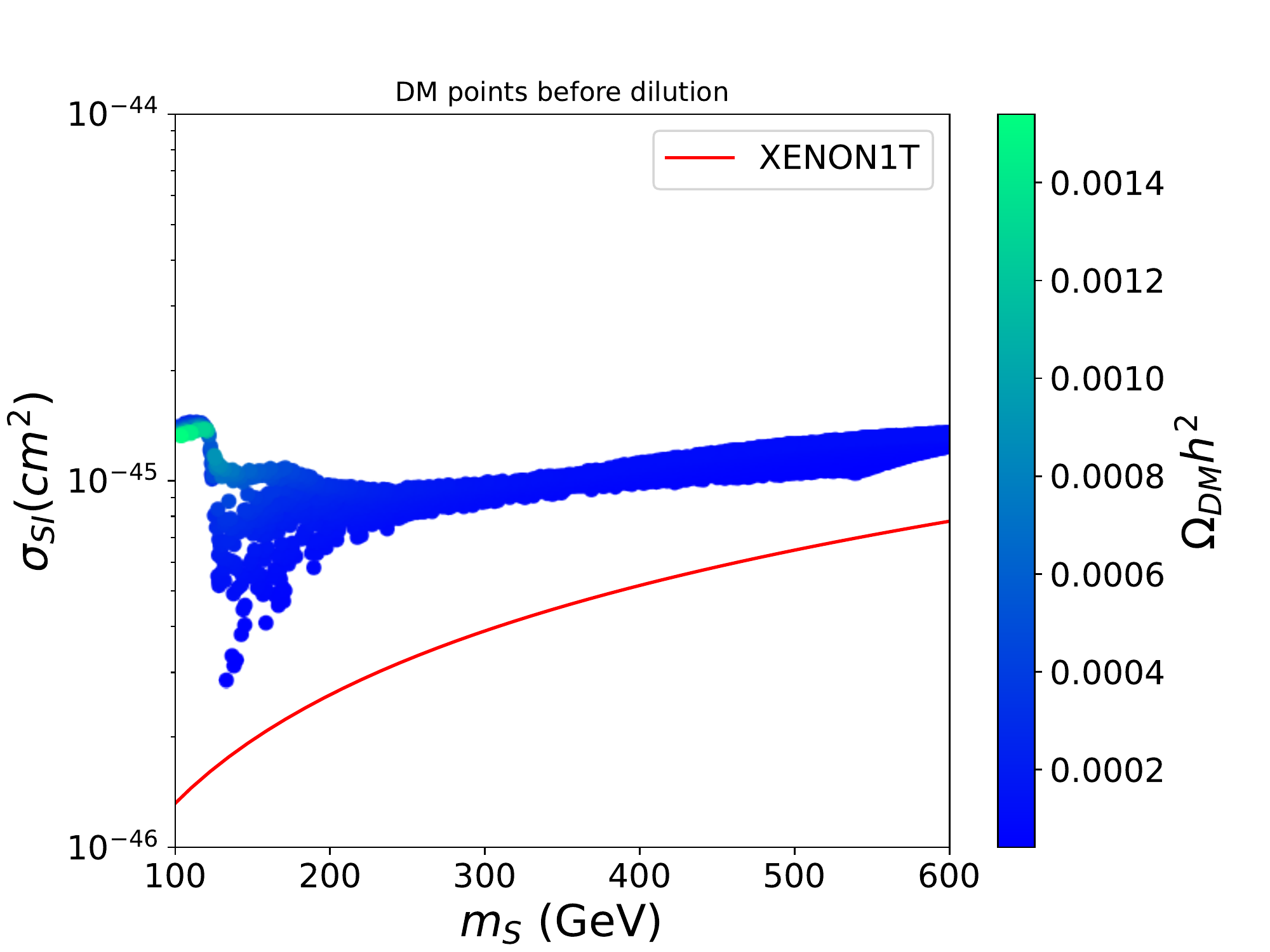}
\hfill
\includegraphics[width=0.48\textwidth]{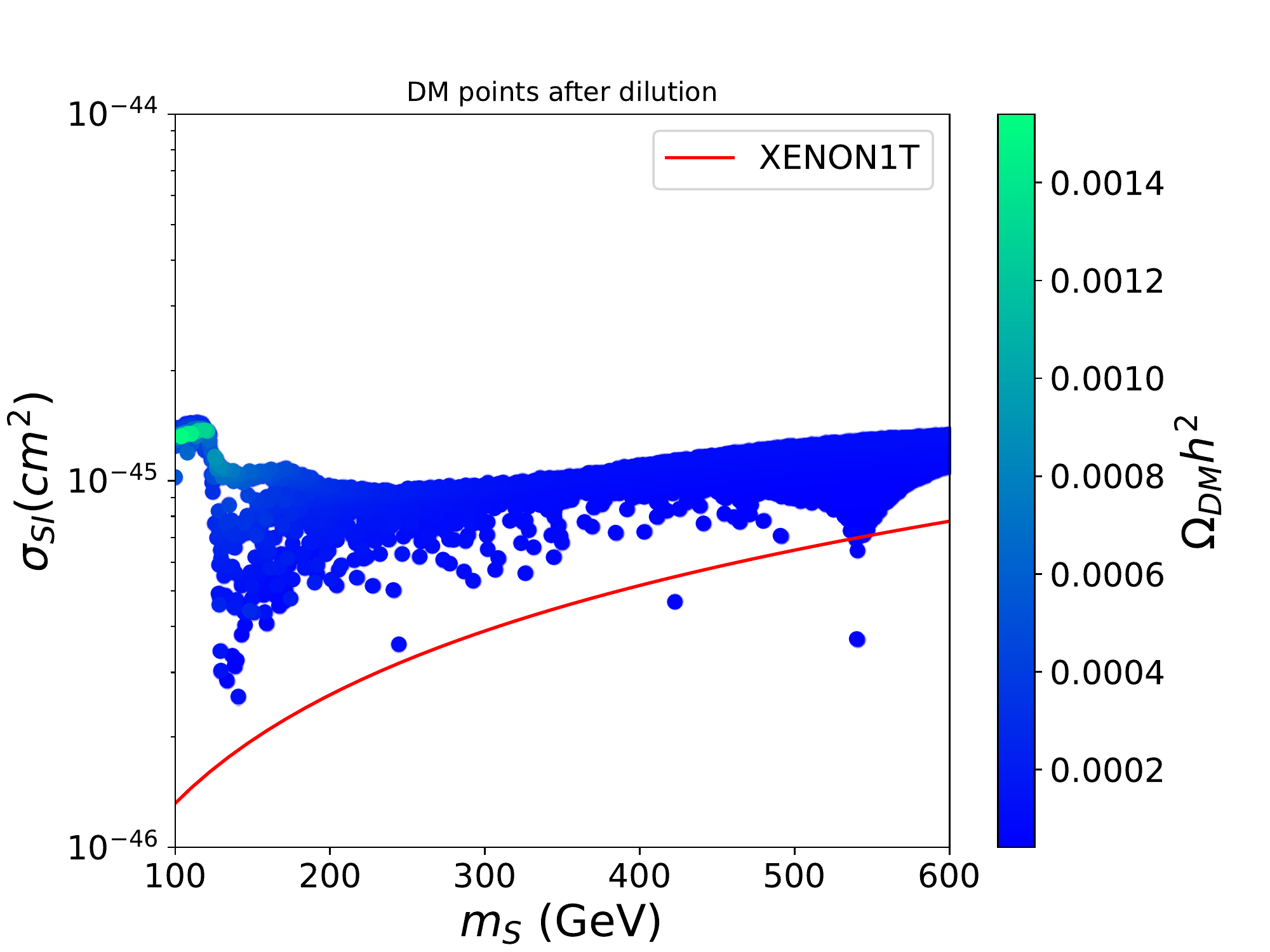}
\caption{{\bf Top:} the dilution factor (left) and supercooling information (right) in the plane of
input parameters. {\bf Bottom:} the cross section of spin-independent scattering of DM on nucleon without (left) and with (right) dilution of DM relic density caused by FOPT taken into consideration, along with the 90\% CL limit from XENON1T.}
\label{fig:dm}
\end{figure}

The top left panel of \cref{fig:dm} displays the results of the dilution factor as a function of the model parameters. 
It finds that the dilution effect can be neglected in most of the parameter space that can achieve FOPT, except at the upper edge on $(m_s,\lambda_{hs})$ plane.
In the upper edge, the phase transition happens mostly between minimums of $(v_h^{\rm high}=0,v_s^{\rm high}\neq 0)$ and $(v_h^{\rm low}\neq 0,v_s^{\rm low}=0)$, which is strong first-order and corresponds to a large difference between $T_c$ and $T_n$. 
It can be seen from the top panels of \cref{fig:dm}, the larger the difference is, the more obvious the dilution is. 
This is consistent with the results in previous studies that the dilution can be neglected when $T_{c} \sim T_{n}$~\cite{bian2018thermally, bian2019two}, and is significant if supercooling occurs~\cite{wainwright2009impact,megevand2008supercooling}. The impact of supercooling can be expected from \cref{eq: supercolling dilution factor}. At high temperatures, the dominant part of effective potential is the quartic term of temperature provided by Eq~(\ref{eq: one-loop effective potential at finite temperature}). Thus the dilution factor can be estimated as  
\begin{equation} \label{eq: dilution approx}
  s = - \frac{{\rm d} V_{\rm eff}}{{\rm d} T}, 
  ~~~~~~~~~~
  (\frac{a_{f}}{a_{i}})^3 \sim (\frac{T_{c}}{T_{n}})^3 \sim (1 + \frac{\Delta}{T_{n}})^3.
\end{equation}

In the bottom panels of \cref{fig:dm}, we present the cross section of spin-independent scattering of DM on nucleon for the above samples, along with the 90\% CL limit from XENON1T~\cite{PhysRevLett.119.181301}. In the case of singlet DM relic density $\Omega_{S} h^2$ smaller than the observed value $\Omega_{\rm DM} h^2 = 0.120$, which means that the singlet DM is only a fraction of DM, we re-scale the cross section by a factor of $f_{\rm rel} = \Omega_{S}/ \Omega_{\rm DM}$. The bottom panels show the results without and with the dilution of $\Omega_{S} h^2$. One can see that the dilution can salvage some samples excluded by DM direct detection. This is the reason why the dilution caused by FOPT should be taken into consideration in DM studies.

\begin{figure}[t]
\centering 
\includegraphics[width=0.48\textwidth]{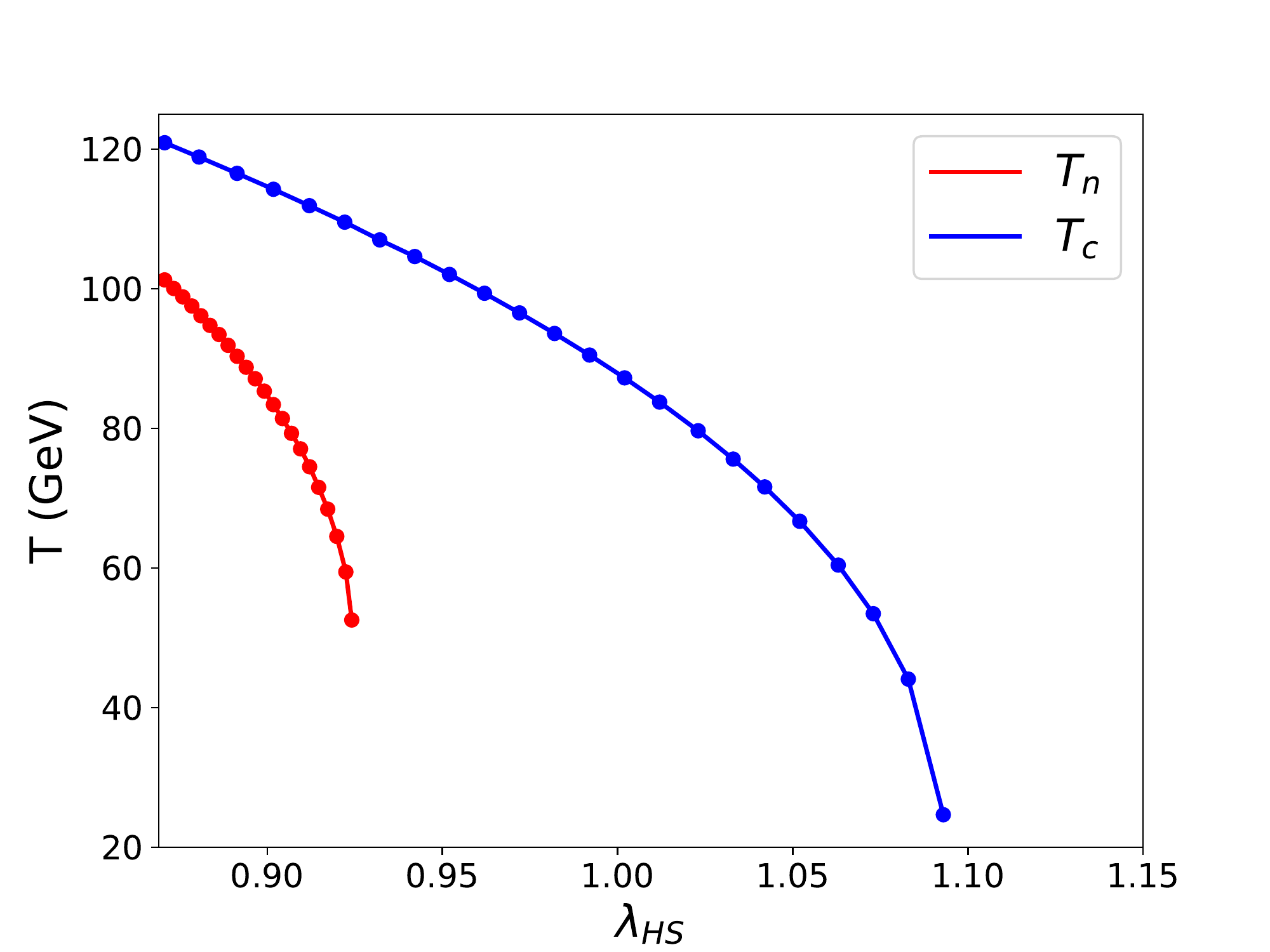}
\hfill
\includegraphics[width=0.48\textwidth]{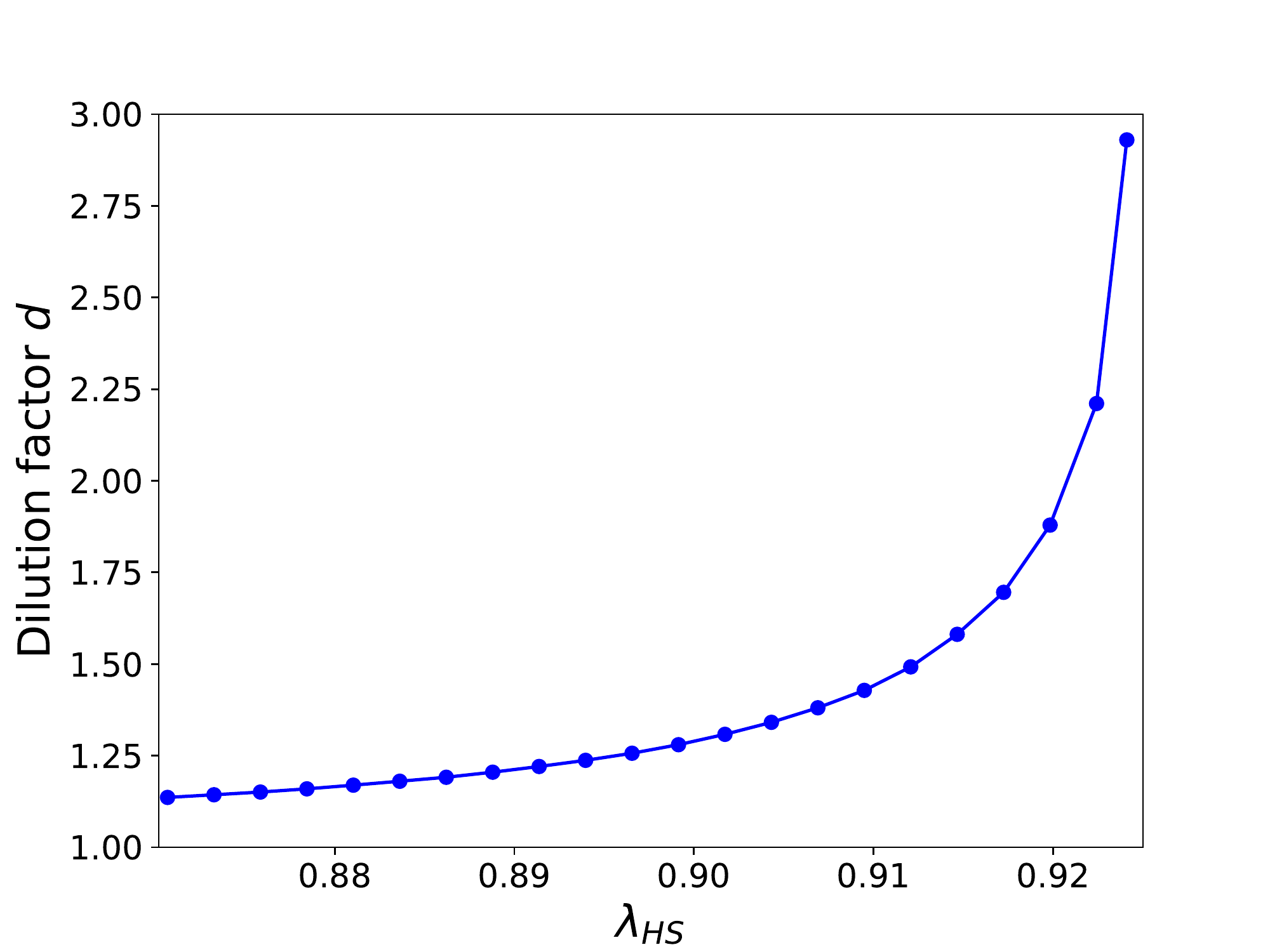}
\caption{The critical temperature $T_c$, the nucleation temperature $T_n$ and the dilution factor $d$ as functions of the input parameter $\lambda_{HS}$, with fixed $m_s = 129.5070$ GeV and $\lambda_{S} = 0.5286$.}
\label{fig:bk1}
\end{figure}
\begin{figure}[t]
\centering 
\includegraphics[width=0.6\textwidth]{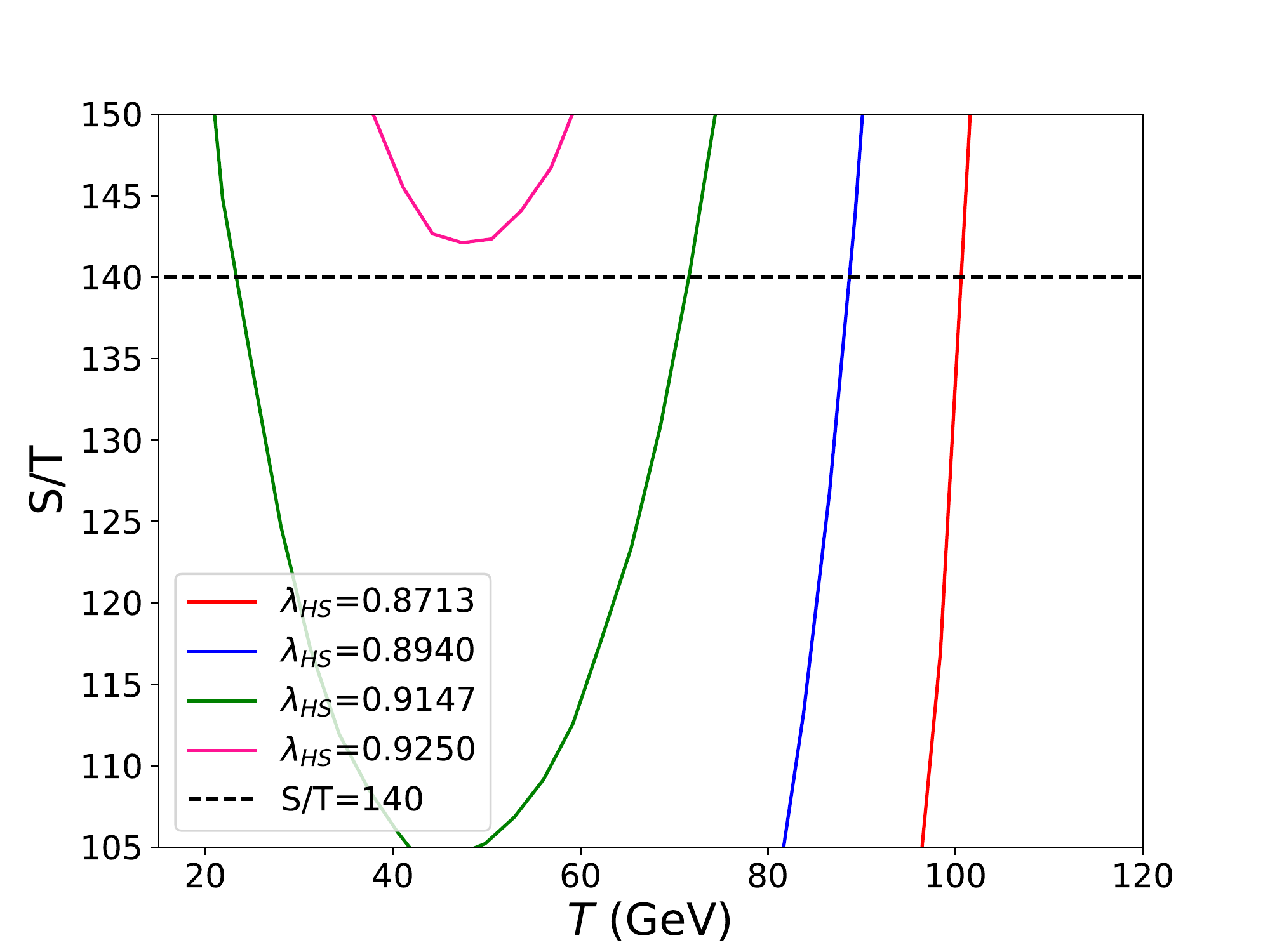}
\caption{The ratio of the Euclidean action to the temperature $S/T$ versus the temperature for different $\lambda_{HS}$ values. Other input parameters are same as in \cref{fig:bk1}.}
\label{fig:bk2}
\end{figure}

The largest value of the dilution factor in our scan is $2.8$, which may be raised a little with more sophisticated scan. However, as it is correlated with the difference $T_c - T_n$, we find that it is hard to obtain  a dilution factor larger than $3$. In \cref{fig:bk1} we displace $T_c$, $T_n$ and the dilution factor $d$ as functions of the model parameter $\lambda_{HS}$ for a benchmark point with $m_s = 129.5070$\gev and $\lambda=0.5286$. 
We see in the left panel that the critical temperature $T_c$ and the nucleation temperature $T_n$ both decrease but with different speeds as $\lambda_{HS}$ increases. Theoretically, the lowest value of $T_c$ can reach to zero by fine tuning the parameter $\lambda_{HS}$, as discussed in Ref.~\cite{Athron:2022jyi}. However, the nucleation temperature has a lower bound around 50\gev. The reason of existing this lower bound is displaced in \cref{fig:bk2}, where the colored solid curves indicate the ratio of the Euclidean action to the temperature as functions of the temperature for different $\lambda_{HS}$ values. The nucleation temperature is obtained from condition $S/T=140$, namely the intersection of the colored curves and the horizontal black dashed line. With $\lambda_{HS}$ increasing, the intersection moves to the left and the slope of $S/T$ decrease. When $\lambda_{HS}>0.9250$, the $S/T$ curve looks U-shaped and there is no more intersection with the horizontal line, which means that the FOPT can not finish in the history of the Universe. 

The difference between $T_{c}$ and $T_{n}$ increases with $\lambda_{HS}$ increasing as well, and stops at about $50\gev$ because of no nucleation temperature for larger $\lambda_{HS}$. Therefore, the dilution factor increases with increasing $\lambda_{HS}$ and has an upper bound about 3 for this benchmark point, as shown in the right panel of \cref{fig:bk1}.

\subsubsection{Constraints on dilution process}

Although the dilution in xSM is sizable to affect DM properties, there are two essential prerequisites in the calculation of dilution factor discribed in \cref{sec:mechanism}. In the case of supercooling the liberated latent heat reheats the system back to a temperature close to $T_c$, and the phase transition happens after the singlet DM freeze-out. Now we check the two conditions for the above scan result.

In \cref{eq: f_define}, the volumetric fraction of the low temperature phase during the reheating stage $f$ must be smaller than 100\%. However, using \cref{eq: f_cal}, it can exceed 100\% especially for samples of large dilution factor. It is because \cref{eq: f_cal} assumes that a large amount transition latent heat brings temperature of the system from $T_n$ back to near $T_c$, which fails when $T_c\gg T_n$. In this case, $T_{n}$ appearing in \cref{eq: supercolling dilution factor} should be replaced by reheating temperature $T_{r}$. The calculation of $T_{r}$ involves processing specific kinetic processes~\cite{megevand2008supercooling}, which is out of the scope of this paper. Neglecting the term including $f$, i.e. setting $f=100\%$, \cref{eq: supercolling dilution factor} approaches to \cref{eq: dilution approx} and gives a larger dilution factor.  Moreover, friction and collisions in the walls of bubbles would also release entropy and increase the dilution~\cite{wainwright2009impact}, which are not included in \cref{eq: supercolling dilution factor}. Thus, breaking of this assumption does not indicates reduction of dilution factor. 

\begin{figure}[t]
\centering 
\includegraphics[width=0.6\textwidth]{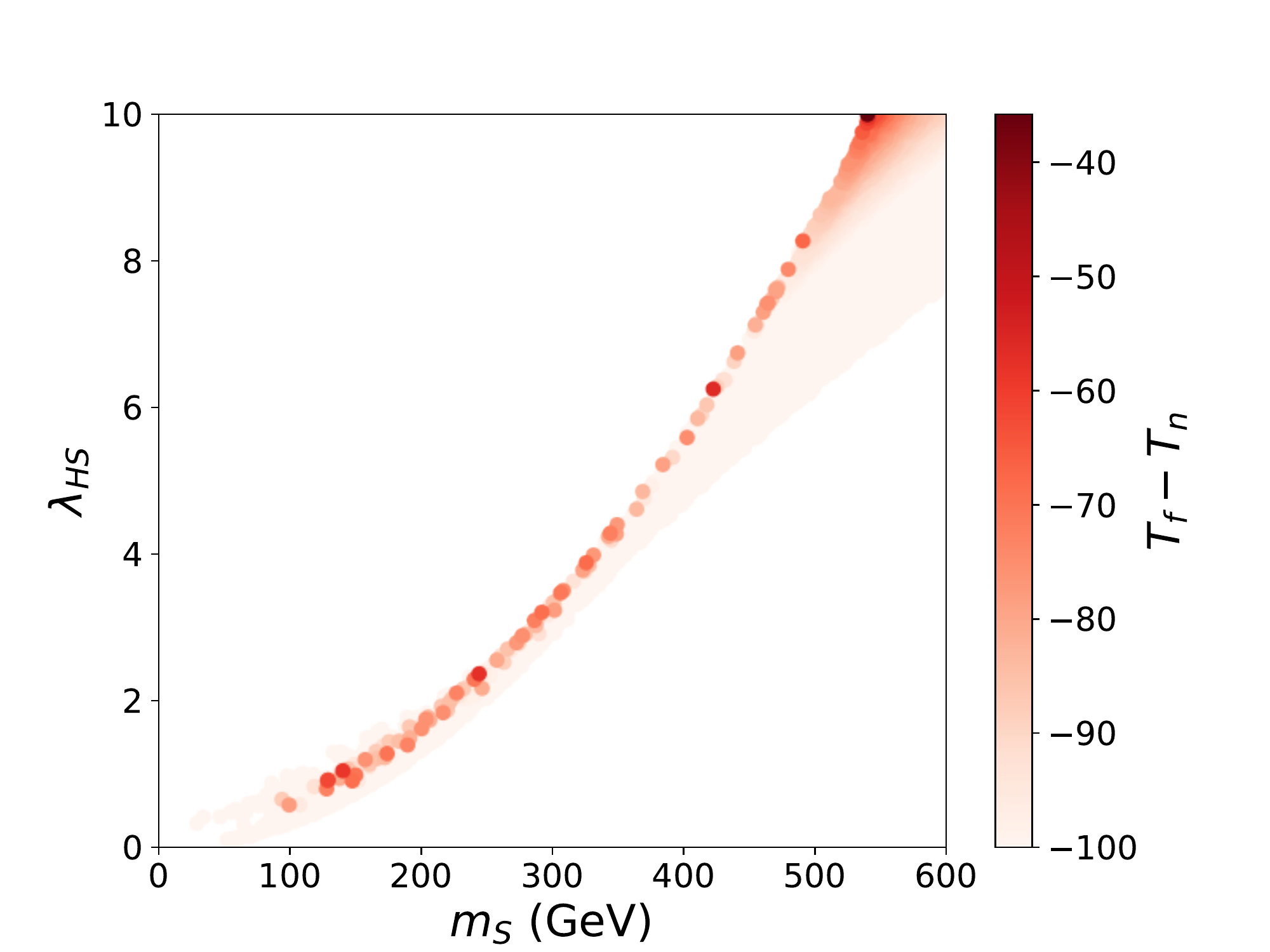}
\caption{The difference between DM freeze-out temperature $T_{f}$ and the nucleation temperature $T_{n}$ in the plane of input parameters. }
\label{fig:diff}
\end{figure}

As for the second prerequisite, in order to avoid the DM reequilibrate with other thermal species, the nucleation temperature $T_{n}$ shall be smaller than the singlet DM freeze-out temperature $T_{f}$. As discussed in \cref{fig:bk2}, there is a lower bound on $T_{n}$, and the lowest $T_{n}$ in our scan is about 50\gev. On the other hand, the variable $x_{f} = m_S/T_f $ used in solving the Boltzmann transport equation of matter lies in range of $\mathcal{O}(20 \sim 40)$ in xSM. We see from \cref{fig:dm} that $m_S$ is smaller than 600\gev when $d>1.5$ and $\lambda_{HS}<10$, which indicates $T_f<30\gev<T_n$, i.e. the dilution process happens before the DM freeze-out and can not affect the current singlet DM density. This can be seen from \cref{fig:diff} which shows that for the samples of FOPT the DM freeze-out temperature is always lower than the nucleation temperature. The mass of singlet DM is bounded because large dilution factor requires transition between minimums $(0,v_s^{\rm high}\neq 0)$ and $(v_h^{\rm low}\neq 0,0)$. The upper bound can be estimated using high temperature approximation.

Using \cref{eq:high-temperature JBF} and neglecting the zero-temperature corrections and the finite temperature corrections beyond $T^2$ terms, we can get the approximate effective potential~\cite{vaskonen2017electroweak}
\begin{equation} \label{eq: V_HT}
    V_{\rm HT}(h,s;T) = \frac{1}{2}(-\mu_{H}^2 + c_HT^2)h^2 + \frac{1}{4} \lambda_{H}h^4 
    +\frac{1}{2}(-\mu_{S}^{2}+ c_ST^2)s^2 + \frac{1}{4} \lambda_{S}s^4
    + \frac{1}{4} \lambda_{HS} h^2s^2,
\end{equation}
where
\begin{equation}
\begin{aligned}
& c_S = \frac{2\lambda_{HS} + 3\lambda_S}{12},\\
& c_H = \frac{9g_{SU(2)_L}^2 + 3g_{U(1)_Y}^2 + 12y_t^2 + 24\lambda_H + 2\lambda_{HS}}{48}.
\end{aligned}
\end{equation}
In order to obtain FOPT between minimums
\begin{equation}
\begin{aligned}
\left(0,~v_s^{\rm high} = \frac{-\mu_{S}^{2}+ c_ST^2}{\lambda_S} \neq 0 \right) 
{\rm ~~~and~~~} 
\left(v_h^{\rm low}=\frac{-\mu_{H}^{2}+ c_HT^2}{\lambda_H} \neq 0 ,~0\right),
\end{aligned}
\end{equation}
with temperature decreasing from high value, $v_s^{\rm high}$ must become non-zero before $v_h^{\rm low}$ become non-zero. It gives 
\begin{equation} \label{eq: contrain_1}
\begin{aligned}
 \frac{\mu_H^4}{c_H^2} < \frac{\mu_S^4}{c_S^2}.
\end{aligned}
\end{equation}

On the other hand, $(v_h^{\rm low},~0)$ must be deeper than $(0, ~v_s^{\rm high})$ when $T<T_c$, such as $T=0$ at zero-temperature minimum, which implies
\begin{equation}  \label{eq: contrain_2}
\begin{aligned}
\frac{\mu_H^4}{\lambda_H} > \frac{\mu_S^4}{\lambda_S}.
\end{aligned}
\end{equation}

Combining the two constraints, we can obtain 
\begin{equation}
\begin{aligned}
\frac{\mu_H^4}{c_H^2}c_S^2 < \mu_S^4 < \frac{\mu_H^4}{\lambda_H}\lambda_S &\rightarrow \frac{\mu_H^2}{c_H}c_S < \mu_S^2 < \frac{\mu_H^2}{\sqrt{\lambda_H}}\sqrt{\lambda_S}.
\end{aligned}
\end{equation}
Note that $\mu_S^2$, $\mu_H^2$ are positive.
Then recalling $\mu_S^2 = \lambda_{HS}\frac{v_{h}^2}{2} - m_{S}^2$, the constrain on $m_{S}$ reads
\begin{equation} \label{eq: ms contrain}
\begin{aligned}
\frac{-\mu_H^2}{\sqrt{\lambda_H}}\sqrt{\lambda_S} + \lambda_{HS}\frac{v_{h}^2}{2} < m_S^2 < \frac{-\mu_H^2}{c_H}c_S + \lambda_{HS}\frac{v_{h}^2}{2}.
\end{aligned}
\end{equation}

\begin{figure}[t]
\centering 
\includegraphics[width=0.48\textwidth]{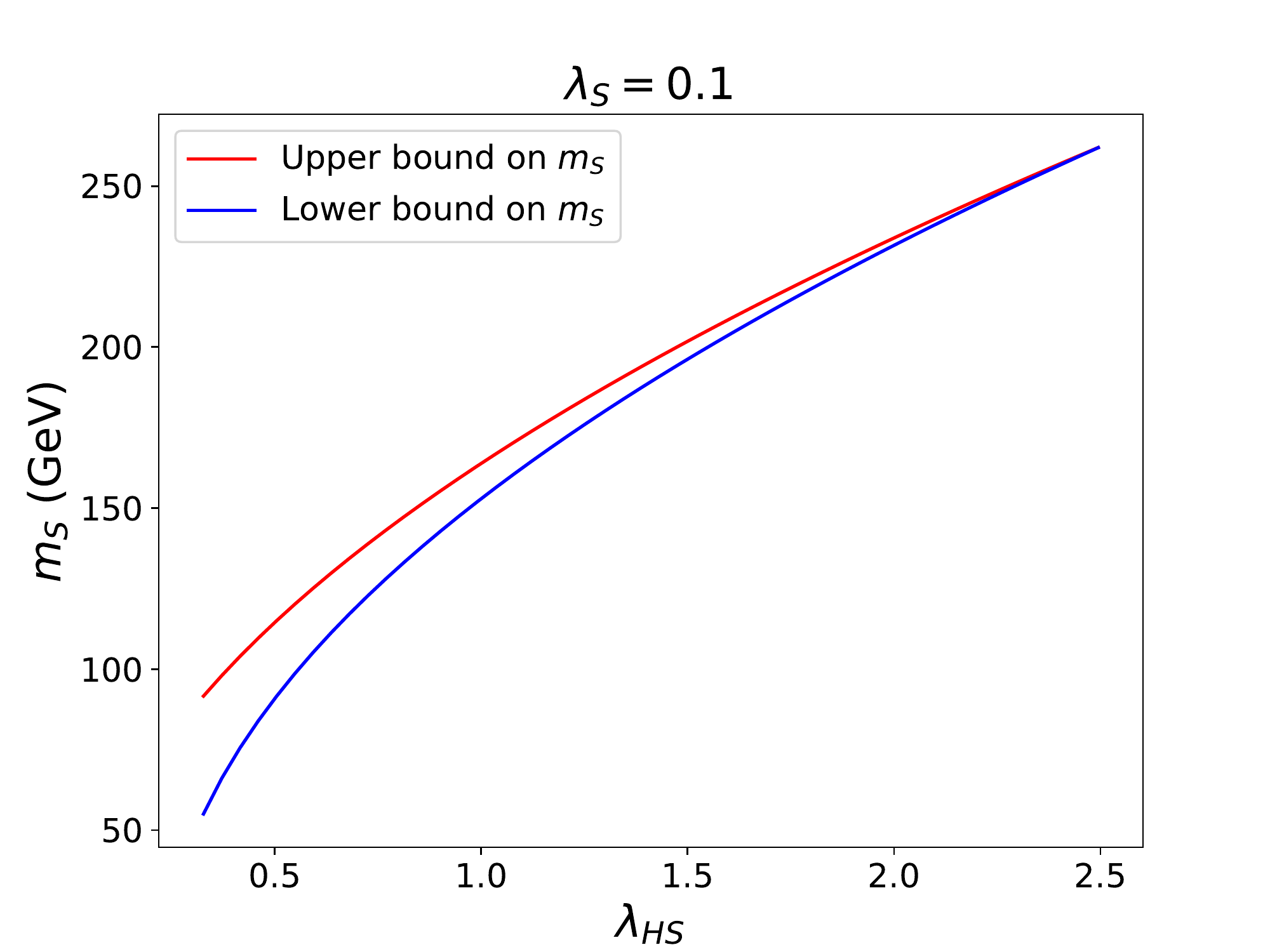}
\includegraphics[width=0.48\textwidth]{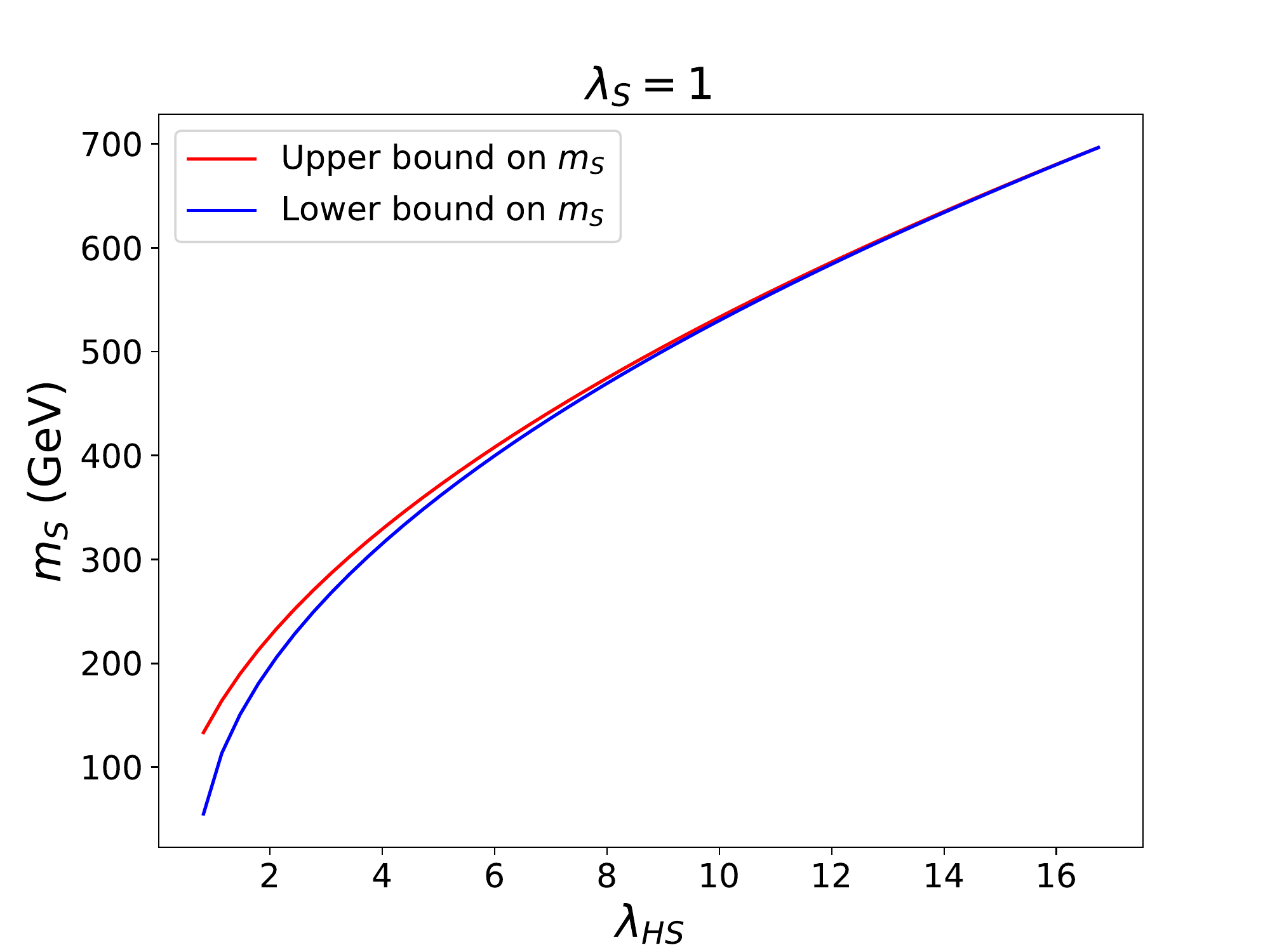}
\caption{The lower and upper bounds on $m_S$ obtained from high-temperature approximated potential for transition between $(0,v_s^{\rm high}\neq 0)$ and $(v_h^{\rm low}\neq 0,0)$.}
\label{fig:mass limit}
\end{figure}

In \cref{fig:mass limit}, we display the upper bound and lower bound on $m_S$ for $\lambda_S=0.1$ and $\lambda_S=1$ with varying $\lambda_{HS}$. The upper bound and lower bound meet in a point, i.e. maximal value of $m_s$. Taking perturbation limits on $\lambda_S$ and $\lambda_{HS}$ into consideration, there is no way that $m_S$ in \cref{eq: ms contrain} exceeds 600\gev. Even without perturbation limits, it is hard to obtain $m_S>1\tev$, namely $T_f>50\gev$. 

By and large, the phase transition always happens after the singlet DM freeze-out in xSM, because the strong FOPT and the perturbation limits set an upper bound of $30\gev$ on $T_f$, while the lowest $T_n$ is above $50\gev$. The dilution of the current singlet DM density in the xSM can be neglected.

\section{Singlet extension of two-Higgs-doublet model}
\label{sec:2HDM+S}

Since the obstacle to achieve the dilution of DM relic density in the xSM is the mass upper bound on the singlet DM, we now turn to the singlet extension of the 2HDM (for a recent review on various 2HDM, see, e.g., \cite{Wang:2022yhm}) to see whether this constraint can be avoided. 

The lightest neutral component pseudoscalar 
$A$ or CP-even state $H$ in the inert 2HDM is stable and can act as a DM candidate, but it is highly restricted. 
The 2HDM itself can achieve a strong electroweak FOPT in the early universe~\cite{Su:2020pjw,Wang:2021ayg}, so the extended real singlet scalar $S$ can be independent of the FOPT.
If so, the VEV of $S$ feld is zero at zero-temperature, and thus $S$  serves as a DM candidate.
For instance, Ref.~\cite{Han:2020ekm} systematically studied the electroweak FOPT and the DM in the type-II 2HDM, taking the relevant constraints into consideration. It showed the surviving parameter space 
where the FOPT happens between the symmetric phase and the broken phase (with $v_s = 0$, and non-zero $v_{h_1}$ and/or $v_{h_2}$), and the singlet field configuration keeps zero in evolution of the Universe. This strongly indicates that the singlet scalar may not be related to the transition temperature. 

We focus on the dilution of DM density in this type-II 2HDM. The tree-level scalar potential is given as 
\begin{equation}
\begin{aligned}
    V_0^{\rm 2HDM+S} &= m_{11}^2(\Phi_{1}^{\dagger}\Phi_{1}) + m_{22}^2(\Phi_{2}^{\dagger}\Phi_{2}) - [m_{12}^2\Phi_{1}^{\dagger}\Phi_{2} + h.c.] \\
    &+ \frac{\lambda_{1}}{2}(\Phi_{1}^{\dagger}\Phi_{1})^2 + \frac{\lambda_{2}}{2}(\Phi_{2}^{\dagger}\Phi_{2})^2 + \frac{\lambda_{3}}{2}(\Phi_{1}^{\dagger}\Phi_{1})(\Phi_{2}^{\dagger}\Phi_{2}) +
    \frac{\lambda_{4}}{2}(\Phi_{1}^{\dagger}\Phi_{2})(\Phi_{2}^{\dagger}\Phi_{1})\\
    &+ [\frac{\lambda_{5}}{2}(\Phi_{1}^{\dagger}\Phi_{2})^2 + h.c.] + \frac{1}{2}S^2(\kappa_1 \Phi_{1}^{\dagger}\Phi_{1}+\kappa_2\Phi_{2}^{\dagger}\Phi_{2}) + \frac{m_{0}}{2}S^2 + \frac{\lambda_{S}}{4!}S^4.
\end{aligned}
\end{equation}
Here we parameterize the two Higgs-doublets in a similar way as in the xSM, assuming all parameters are real, and construct the one-loop effective potential using the OS-like scheme~\cite{Han:2020ekm}. After spontaneous electroweak symmetry breaking, the scalar field $S$ gets a mass 
\begin{equation}\label{eq: 2HDM+S_ms}
    m_{S}^2 = m_{0}^2 + \frac{1}{2}\kappa_{1} v^2 \cos^2\beta + \frac{1}{2}\kappa_{2} v^2 \sin^2\beta,
\end{equation}
where $v=\sqrt{v_{h_1}^2 + v_{h_2}^2} = 246$ GeV. As the DM candidate, its  thermally averaged annihilation cross sections are given by~\cite{Drozd:2014yla}
\begin{equation}
\begin{aligned}
    \left< \sigma_{X\overline{X}} v_{\rm rel}  \right>  &= \sum_{\mathcal{H}\in h,H} \left| \frac{g_{\mathcal{H}SS}C_X^{\mathcal{H}}}{4m_S^2 -m_\mathcal{H}^2 + i\Gamma_\mathcal{H} m_\mathcal{H}} \right|^2 \frac{\Gamma_{\rm SM}({\mathcal{H}^*}\to X\overline{X})}{2m_S} \\
    \left< \sigma_{H_iH_j} v_{\rm rel}  \right> &= \frac{\beta(m_{H_i},m_{H_j})}{32\pi(1+\delta_{ij})m_S^2}\left| g_{H_iH_jSS} 
    + \sum_{\mathcal{H}\in h,H} \frac{g_{\mathcal{H}SS} g_{\mathcal{H} H_iH_j}}{4m_S^2 -m_\mathcal{H}^2 + i\Gamma_\mathcal{H} m_\mathcal{H}}\right.\\
    & \left. ~~~~~~~~~~~~~~ ~~~~~~~~~~~ ~~~~~~~~~~~+2\delta_{\rm CP} \frac{g_{H_i SS}g_{H_j SS}}{\frac{1}{2}(m_{H_i}^2+m_{H_j}^2)-2m_S^2}
    \right|^2,
\end{aligned}
\end{equation}
where $C_X^{\mathcal{H}}$ stands for the normalized coupling of $\mathcal{H}$ to $X\overline{X}$, $\delta_{\rm CP}$ is 0 for $AA$ and $H^+H^-$ and 1 for other cases, and     
\begin{equation}
\beta(m_{H_i},m_{H_j}) = \left(1- \frac{m_{H_i}^2+m_{H_j}^2}{2m_S^2} + \frac{(m_{H_i}^2-m_{H_j}^2)^2}{16m_S^4} \right)^{\frac{1}{2}}.
\end{equation}

The CP-even neutral Higgs couplings can be reorganized as 
\begin{equation}
\begin{aligned}
    \lambda_{h} &= -\kappa_{1}~\sin\alpha ~\cos\beta + \kappa_{2} ~\cos\alpha ~\sin\beta,\\
    \lambda_{H} &= -\kappa_{1}~\cos\alpha ~\cos\beta + \kappa_{2} ~\sin\alpha ~\sin\beta.
    \end{aligned}
\end{equation}
Therefore we can choose the input model parameters as $\tan \beta$, $\sin(\beta-\alpha)$, $m_H$, $m_{H^\pm}$, $m_{12}$, $m_{0}$, $\lambda_h$, and $\lambda_H$. In this way, $m_0$ is a free parameter which is independent of the phase transition properties, so the DM mass $m_{S}$ can be large enough to enhance the freeze-out temperature $T_{f}$ to above the transition temperature $T_{n}$. 

\begin{figure}[t] 
\centering 
\includegraphics[width=0.48\textwidth]{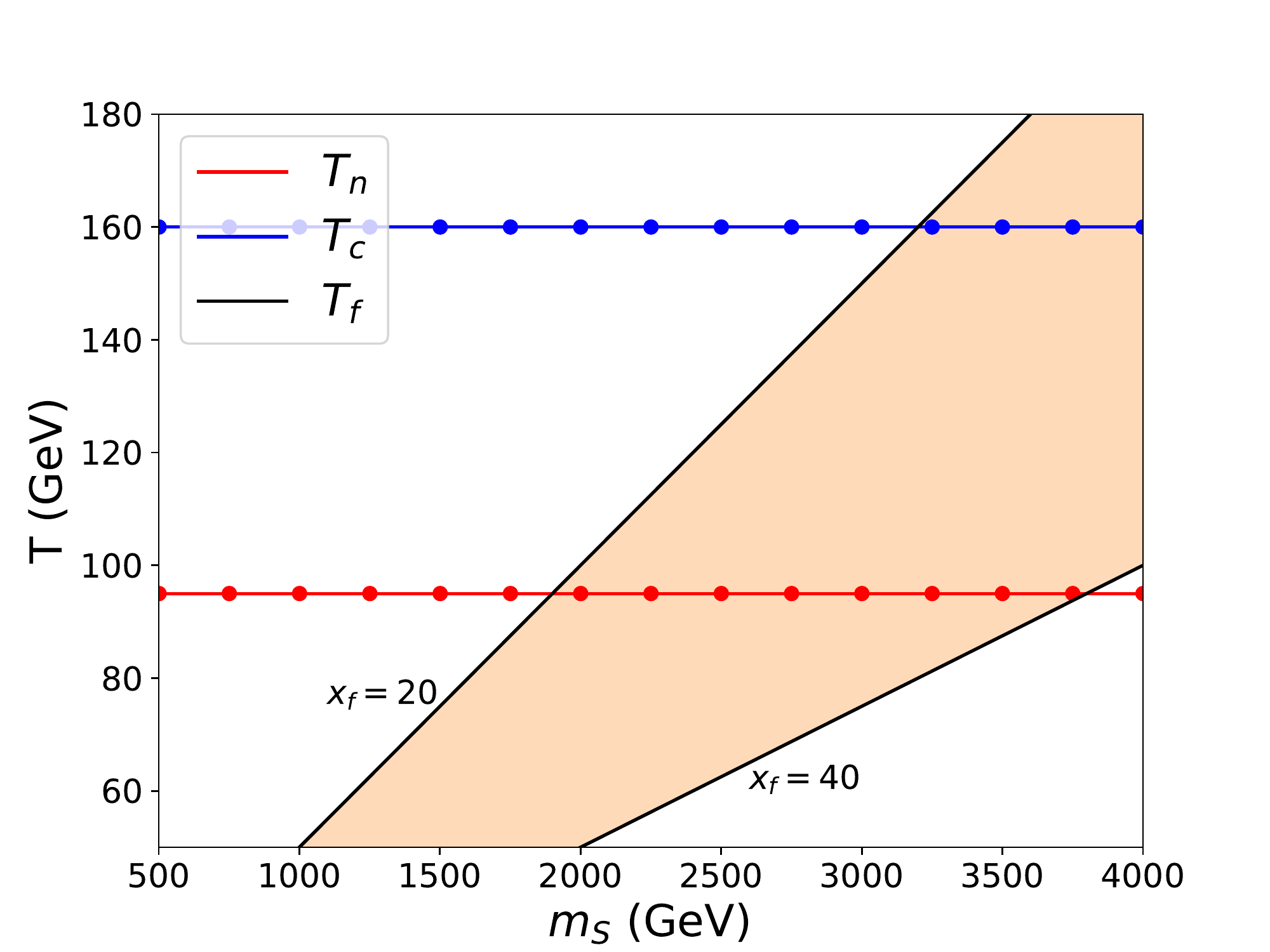}
\includegraphics[width=0.48\textwidth]{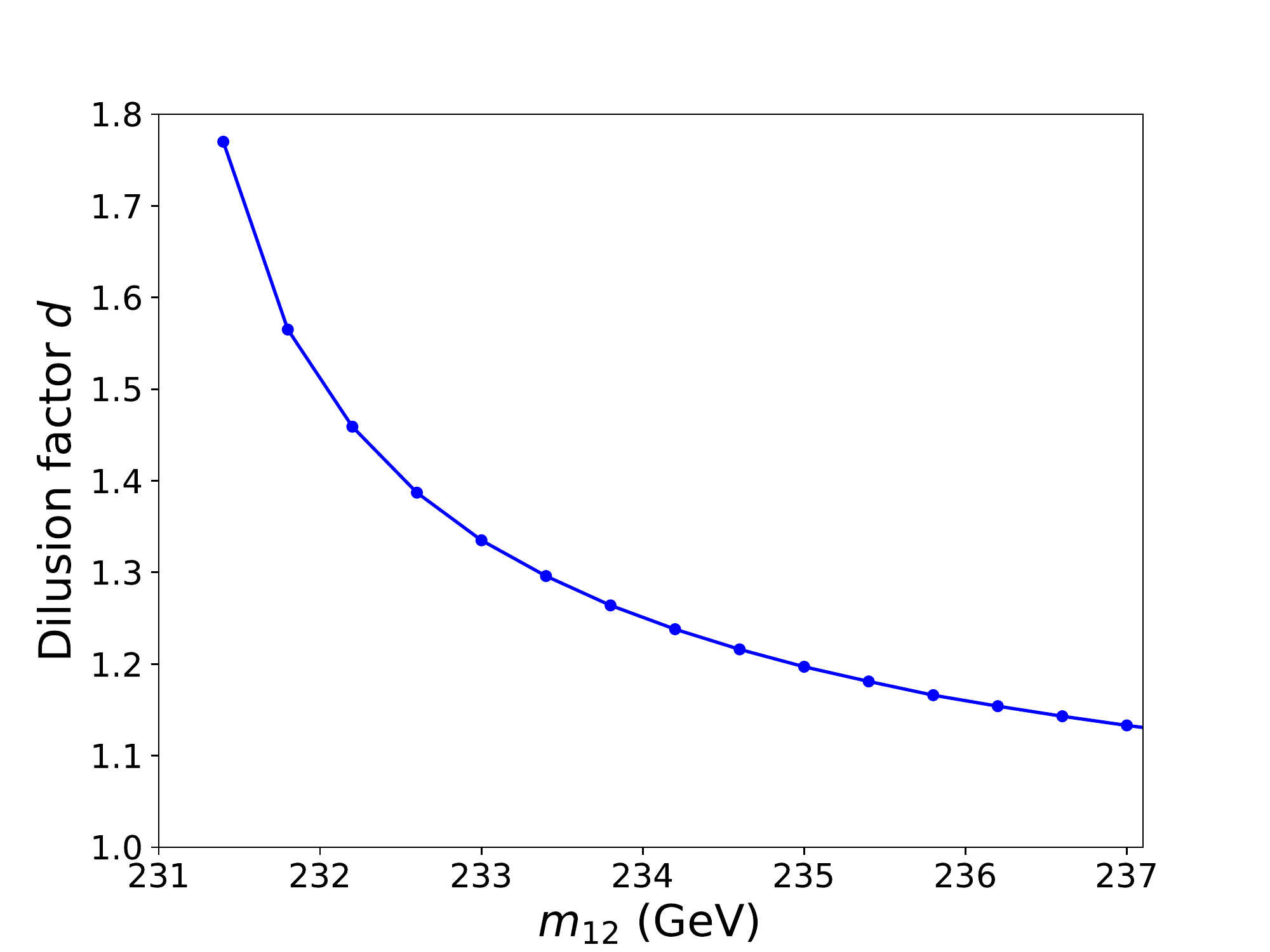}
\caption{The critical temperature $T_c$, the nucleation temperature $T_n$ and the freeze-out temperature $T_f$ as functions of $m_{S}$ (left), and the dilution factor $f$ as a function of $m_{12}$ (right) for the 2HDM+S.}
\label{fig:2HDM+S}
\end{figure}

In \cref{fig:2HDM+S} we display $T_c$, $T_n$ and $T_f$ with a varying $m_0$ in the left panel, and the dilution factor $d$ with a varying mixing parameter $m_{12}$ in the right panel around the benchmark point~\cite{Han:2020ekm}
\begin{equation}
\begin{aligned}
&\tan \beta = 1.87,~~~ \sin(\beta-\alpha)=0.9991,~~~ m_H=387.97\gev,~~~ m_{H^\pm} = 618.31\gev,\\
&m_{12} = 231.41\gev,~~~~ m_S= 501.7\gev,~~~~ \lambda_H=-0.129,~~~~ \lambda_S = 10.93
\end{aligned}
\end{equation}
We see clearly that the transition temperatures $T_c$ and $T_n$ are independent of $m_0$, and there is no upper bound on  $m_s$. The light brown band indicates the possible range of $T_f$ due to the variation of the parameter $x_f=m_S/T_f$ from $20$ to $40$ with the changing DM coupling. When $m_S>2\tev$, $T_c$ becomes lower than $T_f$ in the parameter space for this benchmark point. The right panel shows that the dilution factor increases with $m_{12}$ decreasing, so does the difference between $T_{c}$ and $T_{n}$. The biggest dilution factor value can reach 1.77. If $m_{12}$ further decreases, there is no more $T_{n}$, as shown in \cref{fig:bk1} for the xSM.


Note that when $T_n<T_f$, the value of $T_f$ obtained from \texttt{MicrOmega} is not accurate, as mentioned in \cref{sec:dm}. When the DM deviates from the thermal equilibrium, the electroweak symmetry is still conserved. Thus, the DM mass appearing in the Boltzmann equation should be
\begin{equation}
    m_{S}^2 = m_{0}^2 + [4\kappa_1+4\kappa_2+\lambda_S]\frac{T^2}{24},
\end{equation}
instead of \cref{eq: 2HDM+S_ms}, where the second part is the thermal correction to the singlet in the situation of no mixing between the singlet and Higgs fields. The couplings of DM and mediated Higgs fields are also temperature dependent. For instance, the temperature dependent couplings in the xSM can be seen from \cref{eq: V_HT}. So the temperature dependent cross section indicates that $\left<v\sigma\right>$ is very different from the result without thermal corrections, as well as the $\lambda$. Some dynamical reactions caused by the finite temperature correction may visibly affect the final DM relic density~\cite{Cohen:2008nb,bian2018thermally,Baker:2017zwx,Chao:2020adk}. These adjustments are beyond the scope of  \texttt{MicrOmega}. For our case, we have seen in \cref{fig:Riccati} that the impact on $T_f$ is not dramatic. As far as the freeze-out temperature $T_f$ is still larger than the transition temperature $T_n$, the dilution effect of FOPT will exists with same magnitude shown in the right panel of \cref{fig:2HDM+S}

\section{Next-to-minimal supersymmetric standard model}
\label{sec:nmssm}

In the singlet extension of the 2HDM, the strong FOPT can also be triggered by the broken singlet vacuum expectation value. A special realization of this FOPT case is the next-to-minimal supersymmetric standard model (NMSSM) \cite{Ellwanger:2009dp}, which is a popular scenario of low energy supersymmetry. Despite of null search results of superparticles at the LHC, so far the low energy supersymmetry still remains as a compelling BSM candidate (for recent reviews, see, e.g.,  \cite{Wang:2022rfd,Baer:2020kwz}).  
Besides addressing the baryon asymmetry and the DM issue as well as the hierarchy problem of the electroweak scale~\cite{WITTEN1981267}, the low energy supersymmetry can give a joint explanation for the muon and electron $g-2$ anomalies \cite{Li:2022zap,Li:2021koa,Cao:2021lmj} (it can even marginally  explain the CFD II measurement of the W-boson mass~\cite{Yang:2022gvz,Tang:2022pxh}). 
The NMSSM introduces a SM gauge-singlet chiral superfield to generate the $\mu$-term dynamically and the singlet scalar can couple with the Higgs doublets to sizably  enhance the tree-level mass of the SM-like Higg boson with no need of large radiative effects from heavy top-squarks. So from the point of 125 GeV Higgs boson mass, the NMSSM is more favored than the MSSM (minimal supersymmetric standard model)~\cite{Cao:2012fz}.  

The NMSSM can be treated as a special case of the singlet extension of the 2HDM, where the strong FOPT is triggered by the broken singlet vacuum expectation value and hence relevant to the $\mu$ parameter (the higgsino mass).  

In this model, assuming R-parity symmetry,  the lightest neutralino is usually taken as the DM candidate, which is the mixing of neutral higgsinos, gauginos and singlino. Thus, the DM mass is bounded below by the smallest one among the higgsino mass $\mu$, bino mass $M_1$, wino mass $M_3$ and singlino mass $2\kappa v_S$. The annihilation channels are much more complex than in above models, which are dependent on the component of DM, and there may be also co-annihilations with sleptons or squarks. As studied in \cite{Huang:2014ifa,Bian:2017wfv,Athron:2019teq,Chatterjee:2022pxf,Baum:2020vfl}, 
to achieve a strong FOPT in the NMSSM, the $\mu$ parameter must be smaller than 1\tev . As a result, the neutralino DM must be lighter than 1\tev and the freeze-out temperature $T_f < 50\gev$ is lower than the nucleation temperature as in the xSM.
In conclusion, the strong FOPT happens before the neutralino DM freeze-out, so the dilution of DM relic density can be neglected in the NMSSM.

\section{Conclusions}
\label{sec:conclusions}

We systematically analyzed the dilution of DM relic density caused by the FOPT in the singlet extension models, including the xSM, 2HDM+S and NMSSM. 
We found that in case of supercooling the released entropy can dilute the DM density to 1/3 at most.
However, the singlet field configure in the xSM is relevant to the strength of the phase transition, which sets an upper limit of 600\gev on the singlet DM mass and an upper limit of 30\gev on the singlet DM freeze-out temperature. Meanwhile, the nucleation temperature is larger than 50\gev from our scan. Thus, the strong FOPT happens before the singlet DM freeze-out and the dilution effect is negligible for the current DM density in the xSM.
In the type-II 2HDM+S, the singlet scalar can be independent of the phase transition, so there is no upper bound on the singlet DM freeze-out temperature. As a result, the DM relic density, as well as the DM direct detection, can be affected by the FOPT in the type-II 2HDM+S. 
In the NMSSM with FOPT, the neutralino DM freeze-out temperature is lower than the nucleation temperature as in the xSM and thus the dilution of the DM relic density can be neglected.

\acknowledgments

We would like to thank Subhojit Roy for helpful communications. YZ thanks Institute of Theoretical Physics of Chinese Academy of Sciences (CAS) and YX thanks Zhengzhou University for hospitality during various stages of this work. 
This work was supported by the National Natural Science Foundation of China (NNSFC) under grant numbers 12105248, 11821505 and 12075300,
by Peng-Huan-Wu Theoretical Physics Innovation Center under grant number 12047503,
by a Key R\&D Program of Ministry of Science and Technology of China under grant number 2017YFA0402204,
by the Key Research Program of the Chinese Academy of Sciences under grant number XDPB15,
by the CAS Center for Excellence in Particle Physics (CCEPP),
and by Zhenghou University Young Talent Program.

\addcontentsline{toc}{section}{References}
\bibliographystyle{JHEP}
\bibliography{bibliography}

\end{document}